\documentclass[12pt,a4paper]{article}
\usepackage{latexsym}
\usepackage{amssymb}
\usepackage{amsmath}
\usepackage{amsfonts}
\usepackage{mathrsfs}
\usepackage{cite}
\usepackage{bm}
\usepackage{graphics}
\usepackage{graphicx,array}

\catcode`@=11
\renewcommand{\section}{\@startsection{section}{1}{0pt}{\medskipamount}
{\medskipamount}{\large\bf}} \numberwithin{equation}{section}
\catcode`@=12

\newcommand{\be}{\begin{equation}}
\newcommand{\ee}{\end{equation}}

\def\a{\alpha}

\def\b{\beta}

\def\tr{{\rm tr}}
\def\tr{{\rm tr}\,}
\def\Tr{{\rm Tr}\,}
\def\cN{{\cal N}}

\def\bea{\begin{eqnarray}}
\def\eea{\end{eqnarray}}
\def\nn{\nonumber}

\def\cN{{\cal N}}

\def\f{\frac}

\def\tr{{\rm tr}\,}

\def\nn{\nonumber}

\def\d{\delta}

\def\sB{\stackrel{\frown}{\square}}

\def\eq{\eqref}
\def\pr{\partial}
\def\nb{\nabla}

\numberwithin{equation}{section}

\date{\it  }

\begin{document}

\title{Harmonic superspace approach to the effective action
 in six-dimensional supersymmetric gauge theories}

\author{I.L. Buchbinder\footnote{joseph@tspu.edu.ru}\\
{\small{\em Department of Theoretical Physics, Tomsk State Pedagogical
University,}}\\
{\small{\em 634061, Tomsk,  Russia}} \\
{\small{\em National Research Tomsk State University, 634050, Tomsk, Russia}},\\
{\small{\em Bogoliubov Laboratory of Theoretical Physics, JINR, 141980 Dubna, Moscow region,
Russia}},\\
\\
E.A. Ivanov\footnote{eivanov@theor.jinr.ru}\\
{\small{\em Bogoliubov Laboratory of Theoretical Physics, JINR, 141980 Dubna, Moscow region,
Russia}},\\
\\
B.S. Merzlikin\footnote{merzlikin@tspu.edu.ru}\\
{\small{\em Department of Theoretical Physics, Tomsk State Pedagogical
University}},\\
{\small{\em 634061, Tomsk,  Russia}}, \\
{\small{\em Division of Experimental Physics}},\\
{\small{\em \it National Research Tomsk Polytechnic University, 634050, Tomsk, Russia}},\\
{\small{\em Bogoliubov Laboratory of Theoretical Physics, JINR,
141980 Dubna, Moscow region,
Russia}},\\
\\
K.V. Stepanyantz\footnote{stepan@m9com.ru}\\ {\small{\em Moscow State University}},\\
{\small{\em Faculty of Physics, Department of Theoretical Physics}},\\
{\small{\em 119991, Moscow, Russia}},\\
{\small{\em Bogoliubov Laboratory of Theoretical Physics, JINR, 141980 Dubna, Moscow region,
Russia}} }

\date{}

\maketitle

\begin{abstract}
We review the recent progress in studying the quantum
structure of $6D$, ${\cal N}=(1,0)$ and ${\cal N}=(1,1)$
supersymmetric gauge theories formulated through unconstrained
harmonic superfields. The harmonic superfield approach allows one to
carry out the quantization and calculations of the quantum
corrections in a manifestly  ${\cal N}=(1,0)$ supersymmetric way. The
quantum effective action  is constructed with the help of the
background field method that secures the manifest gauge invariance
of the results. Although the theories under consideration are not
renormalizable,  the extended supersymmetry essentially improves the
ultraviolet behavior of the lowest-order loops. The ${\cal N}=(1,1)$ supersymmetric
Yang--Mills theory turns out to be finite in the one-loop
approximation in the minimal gauge. Also some two-loop divergences are
shown to be absent in this theory. Analysis of the divergences is  performed both in terms of harmonic
supergraphs and by the manifestly gauge covariant superfield
proper-time method. The
finite one-loop leading low-energy effective action is calculated
and analyzed. Also in the abelian case we discuss the gauge
dependence of the quantum corrections and present its precise form
for the one-loop divergent part of the effective action.
\end{abstract}

\begin{center}
\end{center}

\unitlength=1cm

\tableofcontents

\section{Introduction}

The higher-dimensional supersymmetric gauge theories attract the
significant interest due to their remarkable properties in classical
and quantum domains and profound links with string/brane theory. The
various aspects of quantum structure of such theories were intensively investigated for a long time (see, e.g.,
\cite{Howe:1983jm,Howe:2002ui,Bossard:2009sy,Bossard:2009mn,Fradkin:1982kf,Marcus:1983bd,Smilga:2016dpe,Bork:2015zaa}
and references therein). Although these theories are not
renormalizable because of the dimensionful coupling constant
\cite{Gates:1983nr,Buchbinder:1998qv}, it is very interesting to
understand, to which extent a large number of (super)symmetries can
improve the ultraviolet behaviour. It is expected that
supersymmetries sometimes can help cancelling divergences in the lowest
loops, but in higher orders the divergences still appear even
in the maximally extended supersymmetric models
\cite{Marcus:1984ei}. This looks very similar
to what happens in the case of the supergravity theories, but from the technical
point of view the calculations in higher-dimensional gauge theories
are much simpler.

If we wish to understand how the given symmetry improves
the ultraviolet properties of some theory, it is obviously of importance to
use a regularization and the quantization procedure which preserve
this symmetry. For the higher-dimensional supersymmetric Yang--Mills
(SYM) theories with matter it is highly desirable to keep unbroken the gauge
invariance and off-shell supersymmetry. For example, quantizing $4D$,
${\cal N}=1$ supersymmetric theories in superspace, we ensure a
manifest gauge invariance and supersymmetry at all steps of quantum
calculations \cite{Gates:1983nr,Buchbinder:1998qv}. Unfortunately, sometimes it is impossible to quantize a
theory in such a way  that all supersymmetries are off-shell and manifest. For example, $4D$,
${\cal N}=4$ SYM theory cannot be quantized in an ${\cal N}=4$
supersymmetric manner since the manifest ${\cal N}=4$ formulation of
this theory is  yet lacking. However, $4D$, ${\cal N}=2$
supersymmetry can be kept manifest within the harmonic superspace
formalism
\cite{Galperin:1984av,Galperin:1985bj,Galperin:1985va,Galperin:2001uw,Buchbinder:2001wy,Buchbinder:2016wng}.
This approach can be generalized to $6D$ case with ${\cal
N}=(1,0)$ supersymmetry as a manifest symmetry
\cite{Howe:1983fr,Howe:1985ar,Zupnik:1986da,Ivanov:2005qf,Ivanov:2005kz,Buchbinder:2014sna}.
Note that, although $6D$, ${\cal N}=(1,0)$ supersymmetric theories
look very similar to their $4D$, ${\cal N}=2$ counterparts, there is
an essential difference between the two: in the generic case $6D$, ${\cal N}=(1,0)$
theories are anomalous
\cite{Townsend:1983ana,Smilga:2006ax,Kuzenko:2015xiz,Kuzenko:2017xgh}).
However, for the $6D$, ${\cal N}=(1,1)$ theory the anomalies are canceled.
The manifest gauge symmetry is ensured within the
background field method formulated in
harmonic superspace \cite{Buchbinder:2001wy,Buchbinder:1997ya}.

In this paper we briefly review some recent results
\cite{Buchbinder:2016gmc,Buchbinder:2016url,Buchbinder:2017ozh,Buchbinder:2017gbs,Buchbinder:2017xjb,Buchbinder:2018lbd}
concerning the structure of the ultraviolet divergences and low-energy
effective action in $6D$, ${\cal N}=(1,1)$ and ${\cal N}=(1,0)$
SYM theories in the harmonic superspace approach
\footnote{The maximally supersymmetric Yang-Mills theories can be
constructed in a  manifestly supersymmetric way in the pure
spinor superfield formalism \cite{Cederwall:2013vba,Cederwall:2017ruu}. However, the quantum
aspects of this formulation have not been worked  out for the time being, and for this reason we do not
discuss it here.}. The main purpose of this study is to
reveal the structure of the off-shell divergences in the harmonic
superspace approach and to find them explicitly in the lowest loops
following the proposals of Ref. \cite{Bossard:2009mn}. Such
calculations can be done using either the formalism of
harmonic supergraphs, or the harmonic superspace
generalization of the proper time method of Refs. \cite{Schwinger:1951nm,DeWitt:1965jb}.
The proper time method is a
powerful tool of performing the one-loop calculations. In particular,
it well suits for calculating the finite contributions to
effective action in the manifestly gauge invariant and supersymmetric
way. We explicitly  demonstrate the advantages
of the harmonic superspace approach for studying the quantum structure
of $6D$ SYM theories. Though these theories are not renormalizable because of dimensionful coupling constant,
we will see that in the
one-loop approximation ${\cal N}=(1,1)$ SYM theory is finite, if the calculations are performed in
the Feynman gauge.  The absence of divergences in a minimal gauge and
their presence in the non-minimal gauges was already encountered in some
other calculation (see, e.g., \cite{Aleshin:2016yvj}).

The paper has the following structure. In Sect.
\ref{Section_Harmonic_Superspace} we introduce $6D$, ${\cal
N}=(1,0)$ harmonic superspace and explain how it can be used for
formulating supersymmetric gauge theories. Actually, we consider
${\cal N}=(1,0)$ SYM theory interacting with a massless matter hypermultiplet which belongs to an arbitrary
representation of the gauge group. The simplest abelian theory of
this type is investigated in Sect. \ref{Section_Harmonic_SQED} at
the quantum level. First, in Sect.
\ref{Subsection_SQED_Quantization}, we describe the harmonic superspace quantization,
give an account of the Feynman rules, and deduce the Ward identities
encoding the gauge invariance at the quantum level. The next Sect.
\ref{Subsection_One-Loop_Divergences} is devoted to the calculation
of the one-loop divergences and the study of their gauge dependence in the abelian case. In
particular, we construct the total divergent part of the one-loop
effective action and verify that its gauge-dependent part vanishes
on shell. One-loop quantum corrections in the non-abelian case are
investigated in Sect. \ref{Section_Non_Abelian}. We start, in Sect.
\ref{Subsection_One_Loop_Non_Abelian},
with the quantization procedure described in Sect.
\ref{Subsection_Non_Abelian_Quantization} and then calculate the
one-loop divergences, employing the Feynman gauge. In particular, we demonstrate
that in this gauge ${\cal N}=(1,1)$ SYM theory is finite in the one-loop
approximation. The two-loop
divergence of the two-point hypermultiplet Green function (also in the
Feynman gauge) is calculated in Sect.
\ref{Section_Two_Loop_hypermultiplet_Divergences}. We show that for
${\cal N}=(1,1)$ SYM theory this Green function involves no divergences. The calculation of the one-loop divergences by the
harmonic superspace generalization of the proper-time method is
given in Sect. \ref{Manifestly gauge covariant analysis}. This method
is also applied for calculating the finite contributions to the
one-loop effective action in Sect. \ref{Subsection_Finite_Contributions}, where the leading low-energy
structure of this action  was found. It is worth pointing out that such an
effective action is closely related to the on-shell amplitudes in $6D$
maximally extended supersymmetric Yang-Mills theories (see, e.g.,
\cite{Bork:2015zaa} and references therein) and to the so called little
strings \cite{Kutasov:2001uf,Chang:2014jta,Lin:2015zea}.

\section{Harmonic superspace formulation of $6D$ supersymmetric gauge theories}
\label{Section_Harmonic_Superspace}

The conventional $6D, {\cal N}=(1,0)$ superspace is parametrized by the coordinates $z\equiv (x^M, \theta^a_i)$, where $x^M$ with $M=0,\ldots 5$ are
the ordinary space-time coordinates, and $\theta^{a}_i$ with $a=1,\ldots 4$ and $i=1,2$ are the Grassmann ({\it i.e.}, anticommutaing) variables
forming a $6D$ left-handed spinor. The harmonic superspace is obtained from the ${\cal N}=(1,0)$ superspace just defined by adding to its coordinates
the harmonic variables $u^{\pm i}$, such that $u^{+i} u_i^- = 1$ and $u_i^- = (u^{+i})^*$.

The basic novel feature of the harmonic superspace is the existence of an analytic subspace in it, with the coodinates

\begin{equation}
x^M_A \equiv x^M + \frac{i}{2}\theta^{-}\gamma^M \theta^+;\qquad \theta^{\pm a} \equiv u^\pm_i \theta^{ai};\qquad u^{\pm}_i\,.
\end{equation}

\noindent
This subspace is closed under $6D$, ${\cal N}=(1,0)$ supersymmetry transformations.

For the integration measures on the harmonic superspace and its analytic subspace we will use the notation

\begin{equation}\label{Integrations}
\int d^{14}z = \int d^6x\,d^8\theta;\qquad \int d\zeta^{(-4)} \equiv \int d^6x\, d^4\theta^+.
\end{equation}

\noindent
Also we introduce the spinor covariant derivatives

\begin{equation}
D^+_a = u^{+}_i D_{a}^i;\qquad D^-_a = u^{-}_i D_{a}^i,
\end{equation}

\noindent
which satisfy the relation $\{D^+_a, D^{-}_b\} = i(\gamma^M)_{ab}\partial_M$, and define

\begin{equation}
(D^+)^4 = -\frac{1}{24}\varepsilon^{abcd} D_a^+ D_b^+ D_c^+ D_d^+.
\end{equation}

\noindent
The integration measures are related by the useful identity

\begin{equation}\label{Measure_Relation}
\int d^{14}z = \int d\zeta^{(-4)} (D^+)^4.
\end{equation}

\noindent
An important ingredient of the approach is the harmonic derivatives

\begin{equation}
D^{++} = u^{+i} \frac{\partial}{\partial u^{-i}};\qquad D^{--} = u^{-i} \frac{\partial}{\partial u^{+i}};\qquad
D^0 = u^{+i} \frac{\partial}{\partial u^{+i}} - u^{-i} \frac{\partial}{\partial u^{-i}}\,,
\end{equation}

\noindent
which constitute the algebra $SU(2)$,
\begin{equation}
[D^{++}, D^{--}] = D^0\,, \quad [D^0, D^{\pm\pm}]  = \pm  D^{\pm\pm}\,.
\end{equation}
In the analytic basis $(x^M_A, \theta^{\pm a}, u^\pm_i)$ the harmonic derivatives acquire some additional terms, the precise form of which can be found in \cite{Bossard:2015dva}.

The harmonic superspace analog of the gauge field is the analytic superfield $V^{++}(z,u)$ which satisfies the condition

\begin{equation}
D^+_a V^{++} = 0
\end{equation}

\noindent
and is real with respect to the ``tilde'' conjugation, $\widetilde{V^{++}} = V^{++}$. Geometrically, this object is the gauge connection covariantizing
the harmonic deivative $D^{++}$,
\begin{equation}
D^{++} \;\Rightarrow \; \nabla^{++} = D^{++} + i V^{++}\,.
\end{equation}

\noindent
The pure $6D$, ${\cal N}=(1,0)$ SYM theory is described by the harmonic superspace action  \cite{Zupnik:1986da}

\begin{equation}\label{Action_Sypersymmetric_Yang_Mills}
S_{\mbox{\scriptsize SYM}} = \frac{1}{f_0^2} \sum\limits_{n=2}^\infty \frac{(-i)^{n}}{n} \mbox{tr} \int d^{14}z\, du_1 \ldots du_n\,
\frac{V^{++}(z,u_1)\ldots V^{++}(z,u_n)}{(u_1^+ u_2^+) \ldots (u_n^+ u_1^+)}\,.
\end{equation}

\noindent
In this expression $f_0$ is the bare coupling constant. The crucial difference of $6D$ case from the similar $4D$ case
is that the coupling constant $f_0$ in six dimensions is dimensionful, $[f_0] = m^{-1}$. Obviously, this gives rise  to lacking of good renormalization
properties at the quantum level.

In the notation accepted in this paper we will always assume that the gauge superfield in the pure Yang--Mills action (\ref{Action_Sypersymmetric_Yang_Mills})
is decomposed over the generators of the fundamental representation, $V^{++}(z,u) = V^{++A} t^A$. The generators $t^A$ satisfy the conditions

\begin{equation}\label{Fundumental_Representation_Generators}
\mbox{tr}(t^A t^B) = \frac{1}{2}\delta^{AB};\qquad [t^A,t^B] = if^{ABC} t^C,
\end{equation}

\noindent
where $f^{ABC}$ are the gauge group structure constants. Just as in the non-supersymmetric case,  only terms quadratic
in the gauge superfield $V^{++}$ survive in the action (\ref{Action_Sypersymmetric_Yang_Mills}) for the abelian gauge group $G=U(1)$.

General $6D$, ${\cal N}=(1,0)$ gauge theories also involve the hypermultiplets  minimally coupled to the gauge superfield $V^{++}$. In the harmonic superspace
approach the hypermultiplets are described by analytic superfields $q^+$ and their tilde-conjugated $\widetilde q^+$,

\begin{equation}
D^+_a q^{+} = 0;\qquad D^+_a \widetilde q^{+} = 0.
\end{equation}

\noindent
The full action of the gauge theory with hypermultiplets reads

\begin{eqnarray}\label{Action}
S = \frac{1}{f_0^2} \sum\limits_{n=2}^\infty \frac{(-i)^{n}}{n} \mbox{tr} \int d^{14}z\, du_1 \ldots du_n\,
\frac{V^{++}(z,u_1)\ldots V^{++}(z,u_n)}{(u_1^+ u_2^+) \ldots (u_n^+ u_1^+)} - \int d\zeta^{(-4)} du\,\widetilde q^+ \nabla^{++} q^+.
\end{eqnarray}

\noindent
Note that the covariant harmonic derivative in the second piece of this action,

\begin{equation}\label{Covariant_Derivative}
\nabla^{++} = D^{++} + i V^{++} = D^{++} + i V^{++A} T^A\,,
\end{equation}

\noindent
includes the generators $T^A$ corresponding to the representation $R$ to which the hypermultiplet superfields $q^+$ belong.
These generators satisfy the relations analogous to (\ref{Fundumental_Representation_Generators}):

\begin{equation}\label{Auxiliary_Representation_Generators}
\mbox{tr}(T^A T^B) = T(R) \delta^{AB};\qquad [T^A,T^B] = if^{ABC} T^C.
\end{equation}

\noindent
Assuming that the gauge group $G$ is simple, we also define $C_2$ and $C(R)_i{}^j$ as

\begin{equation}
f^{ACD} f^{BCD} = C_2\delta^{AB};\qquad C(R)_i{}^j = (T^A T^A)_i{}^j.
\end{equation}

\noindent
Note that $C(R)_i{}^j$ is proportional to $\delta_i^j$ only for an irreducible representation $R\,$. In particular, for the adjoint
representation of a simple group we have

\begin{equation}\label{Adjoint_Group_Factors}
T(Adj)=C_2;\qquad C(Adj)_i{}^j = C_2 \delta_i^j.
\end{equation}

\noindent If the hypermultiplet belongs to the adjoint
representation, $R=Adj$, the action (\ref{Action}) describes
${\cal N}=(1,1)$ SYM theory which possesses a hidden ${\cal N}=(0,1)$ supersymmetry
in addition to the manifest ${\cal N}=(1,0)$ one. This theory is $6D$ analog of $4D$,
${\cal N}=4$ SYM theory. The $4D$, ${\cal
N}=4$ SYM theory is known to possess unique properties in the quantum domain
since it is a completely finite quantum field theory
\cite{Grisaru:1982zh,Mandelstam:1982cb,Brink:1982pd,Howe:1983sr}. One can expect that
the quantum $6D$, ${\cal N}=(1,1)$ SYM theory possesses some remarkable properties as well.

The general ${\cal N}=(1,0)$ gauge theory described by the action (\ref{Action}) is invariant under the gauge transformations

\begin{equation}\label{Gauge_Transformations}
V^{++} \to  e^{i\lambda} V^{++} e^{-i\lambda}  - i e^{i\lambda} D^{++}e^{-i\lambda};\qquad  q^+ \to  e^{i\lambda} q^+;\qquad \widetilde q^+ \to  \widetilde q^+ e^{-i\lambda}
\end{equation}

\noindent
parametrized by an analytic superfield $\lambda$, such that $\lambda = \lambda^A t^A$ for $V^{++} = V^{++A} t^A$ (in the gauge part of the total action),
and $\lambda = \lambda^A T^A$ for $V^{++} = V^{++} T^A$,\ $q^+$, and $\widetilde q^+$ (in the hypermultiplet part).

Also we will need the non-analytic gauge superfield

\begin{equation}\label{V--Definition}
V^{--}(z,u) \equiv \sum\limits_{n=1}^\infty (-i)^{n+1}\int du_1\,\ldots\,du_n\,\frac{V^{++}(z,u_1) \ldots V^{++}(z,u_n)}{(u^+ u_1^+)(u_1^+u_2^+)\ldots (u_n^+ u^+)},
\end{equation}

\noindent
which covariantizes the harmonic derivative $D^{--}$ and satisfies the ``harmonic flatness condition''

\begin{equation}\label{V--Equation}
D^{++} V^{--} - D^{--} V^{++} +i[V^{++}, V^{--}]=0.
\end{equation}

\noindent
An important object is the analytic superfield strength

\begin{equation}
F^{++} \equiv (D^+)^4 V^{--}, \label{F-Definition}
\end{equation}

\noindent
which obeys the off-shell constraint

\begin{equation}
\nabla^{++} F^{++} = 0,
\end{equation}

\noindent
as a consequence of (\ref{V--Equation}) and the analyticity of $V^{++}$. One more useful quantity is a non-analitic superfield $q^-$ which is defined
by the equation

\begin{equation}\label{Q-_Definition}
q^+ = \nabla^{++} q^- = (D^{++} + i V^{++}) q^-.
\end{equation}

\noindent
The solution of this equation is given by the series

\begin{eqnarray}\label{Q-_Expansion}
&&\hspace*{-8mm} q^- = \int \frac{du_1}{(u^+ u_1^+)} q_1^+ -i \int  \frac{du_1\,du_2}{(u^+ u_1^+)(u_1^+ u_2^+)} V^{++}_1 q_2^+ - \int  \frac{du_1\,du_2\,du_3}{(u^+ u_1^+)(u_1^+ u_2^+)(u_2^+ u_3^+)} V^{++}_1 V^{++}_2 q_3^+ + \ldots\nonumber\\
&&\hspace*{-8mm}  =(-i)^{n-1} \sum\limits_{n=1}^\infty \int du_1 \ldots du_n\, \frac{V^{++}_1 \ldots V^{++}_{n-1}}{(u^+ u_1^+) \ldots (u_{n-1}^+ u_n^+)} q_n^+.
\end{eqnarray}

\noindent
The gauge transformations of the superfields $V^{--}$, $F^{++}$ and $q^-$ defined above are as follows

\begin{equation}
V^{--} \to  e^{i\lambda} V^{--} e^{-i\lambda}  - i e^{i\lambda} D^{--}e^{-i\lambda};\qquad F^{++} \to e^{i\lambda} F^{++} e^{-i\lambda};\qquad q^- \to e^{i\lambda} q^-.
\end{equation}

The simplest particular case of the theory (\ref{Action}) corresponds to the gauge group $U(1)$.
The corresponding abelian gauge theory is $6D$, ${\cal N}=(1,0)$ supersymmetric analog of QED,  and it is  described by the action

\begin{eqnarray}\label{QED_Action}
S = \frac{1}{4f_0^2} \int d^{14}z\,\frac{du_1 du_2}{(u_1^+ u_2^+)^2} V^{++}(z,u_1) V^{++}(z,u_2) - \int d\zeta^{(-4)} du\,\widetilde q^+ \nabla^{++} q^+\,,
\end{eqnarray}

\noindent
with $\nabla^{++} = D^{++} + i V^{++}$. In the abelian case the gauge transformations acquire the form

\begin{equation}\label{Gauge_Transformations_Abelian}
V^{++} \to  V^{++}  - D^{++}\lambda; \qquad V^{--} \to  V^{--}  - D^{--}\lambda; \qquad  q^+ \to  e^{i\lambda} q^+; \qquad F^{++} \to F^{++},
\end{equation}

\noindent
and the expression for $V^{--}$ is considerably simplified,

\begin{equation}\label{V--_Abelian}
V^{--}(z,u) = \int du_1\,\frac{V^{++}(z,u_1)}{(u^+ u_1^+)^2}.
\end{equation}

\section{Quantum corrections in $6D$, ${\cal N}=(1,0)$ supersymmetric electrodynamics}
\label{Section_Harmonic_SQED}

\subsection{Quantization, Feynman rules, and Ward identities in the abelian case}
\label{Subsection_SQED_Quantization}

We will start investigating quantum properties of $6D$, ${\cal N}=(1,0)$ gauge theories in harmonic superspace by considering the simplest abelian theory
with the action (\ref{QED_Action}). The quantization procedure in the abelian case requires fixing the gauge. The harmonic superspace analog
of the well-known $\xi$-gauges in QED is obtained by adding, to the original action, the gauge-fixing term,

\begin{equation}\label{Gauge_Fixing_Term}
S_{\mbox{\scriptsize gf}} = - \frac{1}{4f_0^2\xi_0} \int d^{14}z\, du_1 du_2 \frac{(u_1^- u_2^-)}{(u_1^+ u_2^+)^3} D_1^{++} V^{++}(z,u_1) D_2^{++} V^{++}(z,u_2),
\end{equation}

\noindent
where $\xi_0$ is the gauge parameter. As usual, the normalization was chosen so that the Feynman gauge corresponds to $\xi_0=1$. Taking into account
the absence of the Faddeev--Popov ghosts in the abelian case, the generating functional of the theory under consideration has the form

\begin{equation}\label{Generating_Functional_Abelian}
Z = \exp(iW) = \int DV^{++}\,D\widetilde q^+\, Dq^+\, \exp\Big\{i(S+S_{\mbox{\scriptsize gf}} + S_{\mbox{\scriptsize sources}})\Big\}
\end{equation}

\noindent
(as is well known, $W = -i \ln Z$ is the generating functional for the connected Green functions). In harmonic superspace, the source term  can be written as

\begin{equation}\label{Sources}
\int d\zeta^{(-4)}\,du\, \Big[V^{++} J^{(+2)} + j^{(+3)} q^+ + \widetilde j^{(+3)} \widetilde q^+\Big],
\end{equation}

\noindent
where the analytic superfields $J^{(+2)}$, $j^{(+3)}$ and $\widetilde j^{(+3)}$ are the sources for $V^{++}$, $q^+$, and $\widetilde q^+$, respectively.

The 1PI Green functions are generated by the effective action

\begin{equation}\label{Effective_Action_Definition}
\Gamma = W - S_{\mbox{\scriptsize sources}}\,,
\end{equation}

\noindent
with the sources being expressed in terms of the basic superfields  by the equations

\begin{equation}
V^{++} = \frac{\delta W}{\delta J^{(+2)}};\qquad q^+ = \frac{\delta W}{\delta j^{(+3)}};\qquad \widetilde q^+ = \frac{\delta W}{\delta \widetilde j^{(+3)}}.
\end{equation}

Using the standard technique and starting from the functional (\ref{Generating_Functional_Abelian}), one can  construct the Feynman rules for the considered
theory.\footnote{The detailed analysis of the similar $4D$, ${\cal N}=2$ case has been accomplished  in \cite{Galperin:1985bj,Galperin:1985va}.} Namely, we represent
the total classical action as a sum of the free part $S^{(2)}$ which is quadratic in the involved superfields and the interaction part $S_I$ which
encompasses all terms of the higher orders,

\begin{equation}
S+S_{\mbox{\scriptsize gf}} \equiv S^{(2)} + S_I.
\end{equation}

\noindent
This allows us to write the generating functional in the form

\begin{equation}
Z = \exp\left\{i S_I\Big(V^{++} \to \frac{1}{i}\frac{\delta}{\delta J^{++}},\, \widetilde q^+ \to \frac{1}{i} \frac{\delta}{\delta \widetilde j^{(+3)}},\, q^+ \to \frac{1}{i} \frac{\delta}{\delta j^{(+3)}}\Big)\right\} Z_0,
\end{equation}

\noindent
where the generating functional of the free theory is given by the Gaussian integral

\begin{equation}\label{Z_Free}
Z_0 \equiv \int DV^{++}\,D\widetilde q^+\, Dq^+\, \exp\Big\{i\big(S^{(2)}+ S_{\mbox{\scriptsize sources}}\big)\Big\}.
\end{equation}

\noindent
Then the expression for $S_I$ produces the vertices, while all propagators are encoded in $Z_0$.

For the theory (\ref{QED_Action}), the free part of the action and the interaction term read

\begin{eqnarray}\label{Quadratic_Action}
&& S^{(2)} =  \frac{1}{4f_0^2}\Big(1-\frac{1}{\xi_0}\Big) \int d^{14}z\, du_1 du_2 \frac{1}{(u_1^+ u_2^+)^2} V^{++}(z,u_1) V^{++}(z,u_2) \nonumber\\
&&\qquad\qquad\qquad\qquad\qquad + \frac{1}{4f_0^2\xi_0} \int d\zeta^{(-4)}\, du\, V^{++}(z,u) \partial^2 V^{++}(z,u);\qquad\\
\label{Interaction}
&& S_I = - i\int d\zeta^{(-4)}\, du\, \widetilde q^+V^{++} q^+.
\end{eqnarray}

\begin{figure}[h]
\begin{picture}(0,2)
\put(7,0){\includegraphics[scale=0.15]{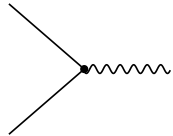}}
\end{picture}
\caption{The only interaction vertex of the abelian $6D$, ${\cal N}=(1,0)$ SQED.}
\label{Figure_Abelian_Vertex}
\end{figure}

\noindent
{}From the interaction (\ref{Interaction}) we conclude that there is only one interaction vertex in the theory. It is depicted in Fig. \ref{Figure_Abelian_Vertex}.

For calculating the Gaussian integral in (\ref{Z_Free}) we solve the free equations of motion (see \cite{Buchbinder:2017ozh} for details)
and substitute the result into the argument of the exponential. This gives

\begin{eqnarray}
&& Z_0 = \exp\left\{ \frac{i}{2} \int d\zeta^{(-4)}_1\, du_1\, d\zeta^{(-4)}_2\, du_2\, J^{++}(z_1, u_1) G_V^{(2,2)}(z_1,u_1;z_2,u_2) J^{++}(z_2,u_2)\right.\qquad\nonumber\\
&&\left. + i \int d\zeta^{(-4)}_1\, du_1\, d\zeta^{(-4)}_2\, du_2\, j^{(+3)}_1 G_q^{(1,1)}(z_1,u_1;z_2,u_2) \widetilde j^{(+3)}_2
\right\}.
\end{eqnarray}

\noindent
Here, the propagators of the gauge superfield and of the hypermultiplet are given, respectively, by the expressions

\begin{eqnarray}\label{Gauge_Propagator}
&& G_V^{(2,2)}(z_1,u_1;z_2,u_2) = - 2 f_0^2 \Big(\frac{\xi_0}{\partial^2} (D_1^+)^4 \delta^{(2,-2)}(u_2,u_1)\nonumber\\
&&\qquad\qquad\qquad\qquad\qquad\qquad\qquad - \frac{\xi_0-1}{\partial^4} (D_1^+)^4 (D_2^+)^4 \frac{1}{(u_1^+ u_2^+)^2}\Big) \delta^{14}(z_1-z_2);\qquad\\
\label{Hypermultiplet_Propagator}
&& G_q^{(1,1)}(z_1,u_1;z_2,u_2) = (D_1^+)^4 (D_2^+)^4 \frac{1}{\partial^2} \delta^{14}(z_1-z_2) \frac{1}{(u_1^+ u_2^+)^3},
\end{eqnarray}

\noindent
with

\begin{equation}
\delta^{14}(z_1-z_2) \equiv \delta^6(x_1-x_2) \delta^8(\theta_1-\theta_2).
\end{equation}

\begin{figure}[h]
\begin{picture}(0,2.7)
\put(5,1.2){\includegraphics[scale=0.2]{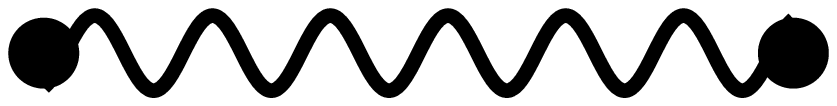}}
\put(4.8,0.7){$V^{++}$} \put(6.4,0.7){$V^{++}$}
\put(9,1.2){\includegraphics[scale=0.2]{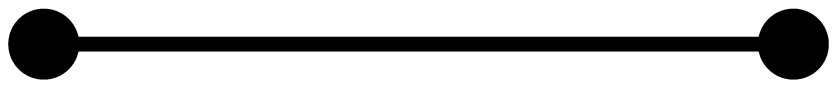}}
\put(8.9,0.7){$\widetilde q^+$} \put(10.5,0.7){$q^+$}
\end{picture}
\caption{The propagators of the gauge superfield $V^{++}$ and the hypermultiplets.}\label{Figure_Propagators_SQED}
\end{figure}

\noindent
Graphically, the $V^{++}$ propagator is denoted by a wavy line, while the hypermultiplet propagator by a solid line.
They are depicted on the left and the right sides of Fig. \ref{Figure_Propagators_SQED}, respectively.

It is obvious that the Feynman diagrams containing closed loops are divergent. Their superficial degree of divergence has been found in Ref. \cite{Buchbinder:2016gmc}.
It is defined by the equation

\begin{equation}\label{Divergence_Degree_QED}
\omega = 2L - N_q - \frac{1}{2} N_D,
\end{equation}

\noindent
where the number of loops is denoted by $L$, the number of external hypermultiplet lines by $N_q$, and $N_D$ denotes the number of spinor covariant derivatives
acting on the external legs. From Eq. (\ref{Divergence_Degree_QED}) one can directly conclude that in the one-loop approximation divergent diagrams should
either contain two external hypermultiplet lines or not contain such external lines at all.

At the quantum level the gauge invariance of the given theory leads to some relations between the Green functions. In the abelian case
these are the Ward identities  \cite{Ward:1950xp}. Their non-abelian generalization is
the Slavnov--Taylor identities \cite{Taylor:1971ff,Slavnov:1972fg}. The harmonic superspace Ward identities were constructed in \cite{Buchbinder:2018lbd}
by making the transformation (\ref{Gauge_Transformations}) in the generating functional (\ref{Generating_Functional_Abelian}). Using the notation

\begin{equation}
\Delta\Gamma = \Gamma - S_{\mbox{\scriptsize gf}},
\end{equation}

\noindent
the generating Ward identity amounts to the equation

\begin{eqnarray}\label{Ward_Identity_Generating}
D^{++}\frac{\delta\Delta\Gamma}{\delta V^{++}} = - i q^+ \frac{\delta\Delta\Gamma}{\delta q^+} + i \widetilde q^+ \frac{\delta\Delta\Gamma}{\delta\widetilde q^+}.
\end{eqnarray}

\noindent
The adjective ``generating'' refers to the fact that in this equation the (super)field arguments are not put equal to zero in advance.
Therefore, Eq. (\ref{Ward_Identity_Generating}) encompasses an infinite set of identities which relate
the longitudinal parts of the $(n+1)$-point Green functions to the $n$-point Green functions.

The lowest-order Ward identity leads to the transversality of quantum corrections to the two-point function of the gauge (super)field.
In the harmonic superspace language it can be obtained by differentiating Eq. (\ref{Ward_Identity_Generating}) twice with respect to $V^{++}$:

\begin{equation}\label{Transversality_QED}
D^{++}_1 \frac{\delta^2\Delta\Gamma}{\delta V^{++}_1 \delta V^{++}_2} = 0,
\end{equation}

\noindent
where the superfield arguments have been set equal to zero at the end.

Similarly, differentiating Eq. (\ref{Ward_Identity_Generating}) with respect to $q^+_2$ and $\widetilde q^+_3$ and again setting the superfields equal to zero afterwards,
we obtain a Ward identity which relates three- and two-point Green functions,

\begin{eqnarray}\label{Usual_Ward_Identity}
&& D^{++}_1 \frac{\delta^3\Delta\Gamma}{\delta V^{++}_1 \delta q^+_2 \delta\widetilde q^+_3} = - i (D_1^+)^4 \delta^{14}(z_1-z_2) \delta^{(-3,3)}(u_1,u_2) \frac{\delta^2\Delta\Gamma}{\delta q^+_1 \delta\widetilde q^+_3}\nonumber\\
&&\qquad\qquad\qquad\qquad\qquad\qquad\qquad + i (D^+_1)^4\delta^{14}(z_1-z_3)\delta^{(-3,3)}(u_1,u_3) \frac{\delta^2\Delta\Gamma}{\delta q^+_2 \delta\widetilde q^+_1}.\qquad
\end{eqnarray}

The Ward identities are a very convenient tool for checking the correctness of various quantum calculations.

\subsection{One-loop divergences and their gauge dependence}
\label{Subsection_One-Loop_Divergences}

According to the relation (\ref{Divergence_Degree_QED}), divergent diagrams should have either $N_q = 0$ or $N_q=2$ of the external hypermultiplet lines
(evidently, odd values of $N_q$ are forbidden).
However, the number of external gauge lines can be arbitrary and the degree of divergence of the diagram is independent of this number.
Nevertheless, the total divergent part of the effective action can be restored by applying to the arguments based on the gauge invariance
encoded in the Ward identities. With this in mind, it is actually enough to calculate the lowest divergent Green functions.

\begin{figure}[h]
\begin{picture}(0,2)
\put(6.5,0){\includegraphics[scale=0.4]{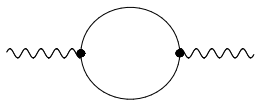}}
\end{picture}
\caption{The supergraph giving the one-loop two-point Green function in the abelian case.}
\label{Figure_Gauge_Diagram}
\end{figure}

For example, the (quadratically divergent) two-point function of the gauge superfield $V^{++}$ in the one-loop order
is determined by the only supergraph  presented in Fig. \ref{Figure_Gauge_Diagram}.

\noindent
Obviously, the expression for it is gauge-independent due to the absence of the gauge propagators. The result obtained in \cite{Buchbinder:2016gmc} can be presented in the form

\begin{equation}\label{Gauge_Part}
\int \frac{d^6p}{(2\pi)^6} \int d^8\theta\, du_1\, du_2\, V^{++}(p,\theta,u_1) V^{++}(-p,\theta,u_2) \frac{1}{(u_1^+ u_2^+)^2} \Big[\frac{1}{4f_0^2} - \frac{i}{2} \int \frac{d^6k}{(2\pi)^6} \frac{1}{k^2 (k+p)^2}\Big].
\end{equation}

When using the dimensional reduction \cite{Siegel:1979wq} to regularize the theory, the divergent part of this expression is

\begin{equation}\label{Gauge_Part_Divergence}
-\frac{1}{6\varepsilon (4\pi)^3}\int d\zeta^{(-4)}\, du\, (F^{++})^2,
\end{equation}

\noindent
where $\varepsilon = 6-D$. However, the regularization by dimensional reduction allows calculating only the logarithmical divergences, while
the considered supergraph diverges quadratically.
For finding these quadratic divergences one needs to use another type of regularization. For example, one could use a special modifications of the Slavnov higher covariant
derivative regularization \cite{Slavnov:1971aw,Slavnov:1972sq} (its harmonic superspace version for $4D$, ${\cal N}=2$ supersymmetric theories was worked
out in \cite{Buchbinder:2015eva}). In the one-loop approximation it suffices to use the simplest ultraviolet cut-off procedure. If the loop momentum
is cut at the scale $\Lambda$, the divergence of the considered contribution to the effective action can be written as \cite{Buchbinder:2017ozh}

\begin{equation}\label{Gamma_QED_CutOff_Abelian}
\int d^{14}z\, du_1\, du_2\, V^{++}(z,u_1) V^{++}(z,u_2) \frac{1}{(u_1^+ u_2^+)^2}
\frac{\Lambda^2}{4(4\pi)^3} - \ln\Lambda \frac{1}{6 (4\pi)^3}\int d\zeta^{(-4)}\, du\, (F^{++})^2.
\end{equation}

\noindent
This expression is gauge invariant, so there appear no further divergent contributions coming from the diagrams with larger numbers of external gauge lines.
Indeed, it is easy to see that the gauge invariant structures proportional to $(F^{++})^n$ with $n\ge 3$ correspond to the finite part of the effective action.

\begin{figure}[h]
\begin{picture}(0,2)
\put(6.5,0){\includegraphics[scale=0.4]{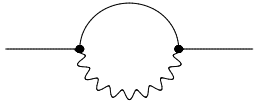}}
\end{picture}
\caption{The supergraph defining the one-loop two-point hypermultiplet Green function.}\label{Figure_Hypermultiplet_2Point}
\end{figure}

Next, let us consider the divergent part of the Green functions with $N_q=2$. The simplest one is the two-point Green function of the hypermultiplet .
In the one-loop order it is given by the logarithmically divergent supergraph presented in Fig. \ref{Figure_Hypermultiplet_2Point}.
The result calculated in \cite{Buchbinder:2018lbd} is given by the gauge-dependent expression

\begin{equation}\label{Hypermultiplet_Green_Function}
- 2if_0^2 \int \frac{d^6p}{(2\pi)^6}\, \frac{d^6k}{(2\pi)^6} \frac{1}{k^4 (k+p)^2} \int d^8\theta\, du_1\, du_2\, \frac{(\xi_0-1)}{(u_1^+ u_2^+)} \widetilde q^+(p,\theta, u_1) q^+(-p,\theta\,u_2),
\end{equation}

\noindent
which is logarithmically divergent in agreement with Eq. (\ref{Divergence_Degree_QED}). The corresponding divergent part (calculated using the  regularization by dimensional reduction)
is written as

\begin{equation}\label{Two_Point_Hypermultiplet_Divergence}
-\frac{2f_0^2}{\varepsilon (4\pi)^3} \int d^{14}z\, du_1\, du_2\, \frac{(\xi_0-1)}{(u_1^+ u_2^+)} \widetilde q^+(z, u_1) q^+(z, u_2).
\end{equation}

\noindent
If applying the cut-off regularization, it is necessary to replace $1/\varepsilon$ by $\ln\Lambda$. We see that the divergence disappears only in the Feynman gauge $\xi_0=1$.

\begin{figure}[h]
\begin{picture}(0,2.5)
\put(6.5,0.){\includegraphics[scale=0.16]{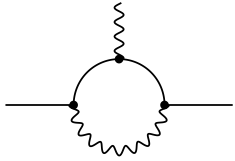}}
\end{picture}
\caption{The harmonic supergraph representing the one-loop contribution to the three-point gauge-hypermultiplet function.}\label{Figure_One-Loop_Vertex_Abelian}
\end{figure}

Surely, the expression (\ref{Two_Point_Hypermultiplet_Divergence}) is not gauge invariant. To obtain the gauge invariant answer, it is necessary to take into
account divergent contributions corresponding to Green functions with $N_q=2$ and an arbitrary number of the external gauge superfield lines. If the number of
the external $V^{++}$ lines is equal to 1, then the corresponding Green function in the one-loop order is contributed to by the only superdiagram
presented in Fig. \ref{Figure_One-Loop_Vertex_Abelian}. The relevant expression was calculated in \cite{Buchbinder:2018lbd}, and it  has the form

\begin{eqnarray}\label{Three-Point_Gauge_Hypermultiplet_Function}
&& 2f_0^2 \int \frac{d^{6}p}{(2\pi)^6}\,\frac{d^{6}q}{(2\pi)^6}\,\frac{d^{6}k}{(2\pi)^6}\,d^8\theta\,\Bigg\{-\int du_1\,du_2\,\widetilde q^+(q+p,\theta,u_1) V^{++}(-p,\theta,u_2) q^+(-q,\theta,u_1)
\nonumber\\
&& \times \frac{\xi_0}{k^2 (q+k)^2 (q+k+p)^2} \frac{1}{(u_1^+ u_2^+)^2}
+ \int du_1\,du_2\,du_3\,\Bigg[ (D^+_{2})^4\,\widetilde q^+(q+p,\theta,u_1)\,V^{++}(-p,\theta, u_2)\nonumber\\
&&\times  q^+(-q,\theta,u_3)\,
\frac{(\xi_0-1)}{k^4 (q+k)^2 (q+k+p)^2}\, \frac{(u_1^+ u_3^+)^2}{(u_1^+ u_2^+)^3 (u_2^+ u_3^+)^3}
- \widetilde q^+(q+p,\theta,u_1)\, V^{++}(-p,\theta, u_2)\nonumber\\
&& \times q^+(-q,\theta,u_3)
\frac{(\xi_0-1)}{k^2 (q+k)^2 (q+k+p)^2} \frac{1}{(u_1^+ u_2^+) (u_2^+ u_3^+)}
- D^+_{2a} D^+_{2b}\,\widetilde q^+(q+p,\theta,u_1)\nonumber\\
&&\times V^{++}(-p,\theta, u_2)\, q^+(-q,\theta,u_3)\,
\frac{(\xi_0-1)(\widetilde\gamma^M)^{ab} k_M}{2 k^4 (q+k)^2 (q+k+p)^2}\, \frac{(u_1^+ u_3^+)}{(u_1^+ u_2^+)^2 (u_2^+ u_3^+)^2}
\Bigg]\Bigg\},
\end{eqnarray}

\noindent
where $(\widetilde\gamma^M)^{ab} = \varepsilon^{abcd} (\gamma^M)_{cd}/2$. It is logarithmically divergent. The divergent part calculated
within the dimensional reduction technique reads \cite{Buchbinder:2018lbd}

\begin{equation}\label{Three_Point_Green_Function_Divergence}
\frac{2if_0^2}{\varepsilon (4\pi)^3} \int d^{14}z \Bigg\{\int du_1\,du_2\,\widetilde q^+_1 V^{++}_2 q^+_1
\frac{\xi_0}{(u_1^+ u_2^+)^2}
+ \int du_1\,du_2\,du_3\, \widetilde q^+_1\, V^{++}_2 q^+_3 \frac{(\xi_0-1)}{(u_1^+ u_2^+) (u_2^+ u_3^+)}
\Bigg\},
\end{equation}

\noindent
where the subscripts denote the harmonic arguments.

To verify the results presented above, it is possible

1. To verify the Ward identity (\ref{Usual_Ward_Identity});

2. To check that the gauge-dependent terms vanish on shell according to the general
theorem of Refs. \cite{DeWitt:1965jb,Boulware:1980av,Voronov:1981rd,Voronov:1982ph,Voronov:1982ur,Lavrov:1986hr}.

\noindent
Both these checks have been done in \cite{Buchbinder:2018lbd}, thereby confirming the correctness of the calculations.

However, so far we have not yet considered all the divergent one-loop diagrams.
Even the sum of the expressions (\ref{Gauge_Part_Divergence}), (\ref{Two_Point_Hypermultiplet_Divergence}) and (\ref{Three_Point_Green_Function_Divergence}),

\begin{eqnarray}\label{Lowest_Gamma_Divergences}
&& \Gamma^{(1)}_\infty = -\frac{1}{6\varepsilon (4\pi)^3}\int d\zeta^{(-4)}\, du\, (F^{++})^2
-\frac{2f_0^2}{\varepsilon (4\pi)^3} \int d^{14}z\, du_1\, du_2\, \frac{(\xi_0-1)}{(u_1^+ u_2^+)} \widetilde q^+_1 q^+_2
+ \frac{2if_0^2}{\varepsilon (4\pi)^3}
\nonumber\\
&& \times \int d^{14}z \Bigg\{\int du_1\,du_2\,\widetilde q^+_1 V^{++}_2 q^+_1
\frac{\xi_0}{(u_1^+ u_2^+)^2}
+ \int du_1\,du_2\,du_3\, \widetilde q^+_1\, V^{++}_2 q^+_3 \frac{(\xi_0-1)}{(u_1^+ u_2^+) (u_2^+ u_3^+)}
\Bigg\}\qquad\nonumber\\
&& + O\Big(\widetilde q^+ (V^{++})^2 q^+\Big),\vphantom{\frac{1}{2}}
\end{eqnarray}

\noindent
is not gauge invariant. The full gauge invariant result can be restored, without further calculations, solely on the ground of gauge invariance considerations.
Below we will show that in the hypermultiplet sector the gauge invariant result is given by an infinite series in $V^{++}$. The expression (\ref{Lowest_Gamma_Divergences})
 is merely a sum of the lowest terms in the $V^{++}$ expansion of the full gauge invariant expression.

 In order to construct the gauge invariant expression for the one-loop divergences we recall the $V^{++}$ series representation (\ref{Q-_Expansion}) for the non-analytic
 superfield $q^-$ defined in (\ref{Q-_Definition}). The first terms of this series read

\begin{equation}\label{Q-_Expansion}
q^- = \int \frac{du_1}{(u^+ u_1^+)} q_1^+ -i \int  \frac{du_1\,du_2}{(u^+ u_1^+)(u_1^+ u_2^+)} V^{++}_1 q_2^+ -\ldots\,.
\end{equation}

\noindent
This representation implies that the total one-loop divergences for $6D$, ${\cal N}=(1,1)$ supersymmetric
electrodynamics in the general $\xi_0$-gauge are written in the form

\begin{eqnarray}\label{One-Loop_Divergence}
&& \Gamma^{(1)}_\infty = -\frac{1}{6\varepsilon (4\pi)^3}\int d\zeta^{(-4)}\, du\, (F^{++})^2 + \frac{2if_0^2 \xi_0}{\varepsilon (4\pi)^3} \int d\zeta^{(-4)}\, du\, \widetilde q^+ F^{++} q^+\nonumber\\
&& - \frac{2 f_0^2 (\xi_0-1)}{\varepsilon (4\pi)^3} \int d^{14}z\, du\, \widetilde q^+\, q^-,
\end{eqnarray}

\noindent
where we also made use of the definition (\ref{F-Definition}) and the precise form (\ref{V--_Abelian}) of $V^{--}$ in the abelian case.

Note that (in agreement with the general theorems \cite{DeWitt:1965jb,Boulware:1980av,Voronov:1981rd,Voronov:1982ph,Voronov:1982ur,Lavrov:1986hr})
the effective action appears to be gauge independent on shell. To demonstrate this, we make use of the on-shell property

\begin{equation}\label{On_Shell_Q-}
q^- = \nabla^{--} q^+,
\end{equation}

\noindent
whence

\begin{eqnarray}
\int d^{14}z\, du\, \widetilde q^+\, q^- = \int d\zeta^{(-4)}\, du\, (D^+)^4\Big(\widetilde q^+\, \nabla^{--} q^+\Big) = i \int d\zeta^{(-4)}\, du\, \widetilde q^+\, F^{++} q^+.
\end{eqnarray}

\noindent
Using this relation, we conclude that all $\xi_0$-dependent terms in the expression (\ref{One-Loop_Divergence}) disappear,

\begin{equation}
\Gamma^{(1)}_\infty\Big|_{\mbox{\scriptsize on shell}} = -\frac{1}{6\varepsilon (4\pi)^3}\int d\zeta^{(-4)}\, du\, (F^{++})^2 + \frac{2if_0^2}{\varepsilon (4\pi)^3} \int d\zeta^{(-4)}\, du\, \widetilde q^+ F^{++} q^+.
\end{equation}

\section{Quantum corrections in non-abelian $6D$, ${\cal N}=(1,0)$ and ${\cal N}=(1,1)$ supersymmetric theories}
\label{Section_Non_Abelian}

\subsection{Quantization of non-abelian $6D$ gauge theories in harmonic superspace by the background field method}
\label{Subsection_Non_Abelian_Quantization}

Let us proceed to investigating the non-abelian case. There are two main differences of the quantization procedure in this case as compared to the abelian one:

1. It is convenient to use the background (super)field method for constructing the manifestly gauge invariant effective action;

2. The gauge-fixing procedure requires adding ghosts.

According to the background field method, we split the gauge (super)field into the background and quantum parts,
so that the theory becomes invariant under two types of gauge transformations. Namely, the background gauge invariance remains unbroken
and so is still a manifest symmetry of the effective action. On the contrary, the quantum gauge invariance is broken by gauge fixing,
although its remnant, the so called BRST symmetry \cite{Becchi:1974md,Tyutin:1975qk}, survives as a symmetry of the total gauge-fixed action.

Within the harmonic superspace formalism the background-quantum splitting is linear. The original superfield $V^{++}$ is presented as a sum
of the background gauge superfield $\bm{V}^{++}$ and the quantum gauge superfield $v^{++}$,

\begin{equation}\label{Splitting_Background_Quantum}
V^{++} = \bm{V}^{++} + v^{++}\,.
\end{equation}

\noindent
The background gauge superfield is treated as an external superfield, for which reason it can appear only on the external legs. We denote the external legs
corresponding to $\bm{V}^{++}$ by the bold wavy lines. The internal and external legs of the quantum gauge superfield will be denoted by the standard wavy lines.

The background-quantum splitting for the hypermultiplets is also possible, but not necessary. The point is that the gauge-fixing term is chosen to be independent of
the hypermultiplet superfields, so the effective action depends only on a sum of the quantum and background hypermultiplet superfields. For this reason here
we do not split the hypermultiplets into the background and quantum parts.

After the background-quantum splitting (\ref{Splitting_Background_Quantum}), the gauge invariance (\ref{Gauge_Transformations}) produces the background gauge invariance

\begin{equation}\label{Background_Gauge_Transformations}
\bm{V}^{++} \to  e^{i\lambda} \bm{V}^{++} e^{-i\lambda}  - i e^{i\lambda} D^{++}e^{-i\lambda}; \qquad v^{++} \to e^{i\lambda} v^{++} e^{-i\lambda}\qquad
q^+ \to  e^{i\lambda} q^+
\end{equation}

\noindent
and the quantum gauge invariance

\begin{equation}\label{Quantum_Gauge_Transformations}
\bm{V}^{++} \to  e^{i\lambda}\bm{V}^{++} e^{-i\lambda}; \qquad v^{++} \to e^{i\lambda} v^{++} e^{-i\lambda}
- i e^{i\lambda} D^{++}e^{-i\lambda};\qquad  q^+ \to  e^{i\lambda} q^+.
\end{equation}

\noindent
Clearly, if we wish to preserve the background gauge invariance as a manifest symmetry of the effective action, it is necessary
to arrange the gauge-fixing term to be invariant under the background transformations. To construct such a term, we introduce the background
bridge superfield related to the superfields $\bm{V}^{++}$ and $\bm{V}^{--}$ as

\begin{equation}
\bm{V}^{++} = - i e^{i\bm{b}} D^{++} e^{-i\bm{b}};\qquad \bm{V}^{--} = - i e^{i\bm{b}} D^{--} e^{-i\bm{b}}.
\end{equation}

\noindent
Then the background gauge transformations (\ref{Background_Gauge_Transformations}) should be supplemented by the transformation of the bridge superfield

\begin{equation}
e^{i\bm{b}} \to e^{i\lambda} e^{i\bm{b}} e^{i\tau},
\end{equation}

\noindent
where a new gauge parameter $\tau=\tau(x,\theta)$ does not depend on the harmonic variables. With the help of the bridge superfield the background gauge invariant
gauge-fixing term is constructed as

\begin{eqnarray}\label{Gauge_Fixing_Term_Non-Abelian}
&& S_{\mbox{\scriptsize gf}} = - \frac{1}{2f_0^2\xi_0}\mbox{tr}\int d^{14}z\, du_1 du_2 \frac{(u_1^- u_2^-)}{(u_1^+ u_2^+)^3} D_1^{++} \Big[e^{-i\bm{b}(z,u_1)}
v^{++}(z,u_1) e^{i\bm{b}(z,u_1)}\Big] \nonumber\\
&& \times D_2^{++} \Big[e^{-i\bm{b}(z,u_2)}v^{++}(z,u_2)e^{i\bm{b}(z,u_2)}\Big].
\end{eqnarray}

\noindent
It is analogous to the usual $\xi$-gauge fixing term for non-supersymmetric Yang--Mills theory, the Feynman (minimal) gauge corresponding to the choice $\xi_0=1\,$.
Note that in the abelian case the dependence on the bridge superfield in (\ref{Gauge_Fixing_Term_Non-Abelian}) is canceled out, and for $6D$, ${\cal N}=(1,0)$
electrodynamics we recover the expression (\ref{Gauge_Fixing_Term}).

As is well known, for quantizing non-abelian theories one should introduce the Faddeev--Popov ghosts. In the background superfield method
the Nielsen--Kallosh ghosts are also needed. In the harmonic superspace language, the Faddeev--Popov ghost action is written as

\begin{equation}\label{Action_Faddeev_Popov_Ghosts}
S_{\mbox{\scriptsize FP}} = \mbox{tr} \int d\zeta^{(-4)}\, du\, b \bm{\nabla}^{++}\Big( \bm{\nabla}^{++} c + i[v^{++},c]\Big).
\end{equation}

\noindent
Here the ghosts $c$ and the antighosts $b$ are the Grassmann analytic superfields in the adjoint representation of the gauge group.
Correspondingly, the background covariant derivative of the ghost superfield takes the form $\bm{\nabla}^{++} c = D^{++} c + i [\bm{V}^{++}, c]$.

In the background superfield method the functional integral after quantization includes determinants which are usually written as
functional integrals over the Nielsen--Kallosh ghosts. Within the harmonic superspace approach such determinants are given by the expression

\begin{equation}\label{Determinant_Nielsev_Kallosh}
\Delta_{NK} \equiv \mbox{Det}^{1/2} \stackrel{\bm{\frown}}{\bm{\Box}} \int D\varphi \exp\big(i S_{\mbox{\scriptsize NK}}\big).
\end{equation}

\noindent
Here we introduced the notation $\stackrel{\bm{\frown}}{\bm{\Box}}\equiv \frac{1}{2} (D^+)^4 (\bm{\nabla}^{--})^2$ and

\begin{equation}
S_{\mbox{\scriptsize NK}} = -\frac{1}{2} \mbox{tr} \int
d\zeta^{(-4)}\, du\, (\bm{\nabla}^{++}\varphi)^2,
\label{Action_Nielsen-Kallosh_Ghosts}
\end{equation}

\noindent
where $\varphi$ are the commuting Nielsen--Kallosh ghosts, analytic Grassmann-even superfields in the adjoint representation.
The determinant $\mbox{Det} \stackrel{\bm{\frown}}{\bm{\Box}}$ in (\ref{Determinant_Nielsev_Kallosh}) can also be cast in the form of a functional integral
by introducing the  Grassmann-odd analytic superfields $\xi^{(+4)}$ and $\sigma$ in the adjoint representation,

\begin{equation}\label{Determinant_Remaining}
\mbox{Det} \stackrel{\bm{\frown}}{\bm{\Box}} = \int D\xi^{(+4)} D\sigma \exp\Big(i\,\mbox{tr}\int d\zeta^{(-4)}\,du\,\xi^{(+4)}
\stackrel{\bm{\frown}}{\bm{\Box}} \sigma \Big).
\end{equation}

Finally, the total generating functional of the theory under consideration takes the form

\begin{equation}
Z = e^{iW} = \int Dv^{++}\,D\widetilde q^+\, Dq^+\,Db\,Dc\,D\varphi\,\mbox{Det}^{1/2} \stackrel{\bm{\frown}}{\bm{\Box}} \exp\Big[i(S+S_{\mbox{\scriptsize gf}}
+S_{\mbox{\scriptsize FP}} + S_{\mbox{\scriptsize NK}}+ S_{\mbox{\scriptsize sources}})\Big].
\end{equation}

\noindent
The sources for the gauge and hypermultiplet superfields differ from the abelian case basically by the presence of the internal symmetry indices,

\begin{equation}\label{Sources_Non-Abelian}
S_{\mbox{\scriptsize sources}} = \int d\zeta^{(-4)}\,du\, \Big[v^{++ A} J^{(+2)A} + j^{(+3)i} (q^+)_i + \widetilde j^{(+3)}_i (\widetilde q^+)^i\Big].
\end{equation}

\noindent
It is necessary to take into account that only the quantum gauge superfield $v^{++}$ is present in the term (\ref{Sources_Non-Abelian}).
In principle, if necessary, it is also possible to introduce sources for ghosts.

The propagators of the quantum gauge superfield and those of the hypermultiplet are similar to those in the abelian case:

\begin{eqnarray}\label{Gauge_Propagator_Non_Abelian}
&& (G_V^{(2,2)})^{AB}(z_1,u_1;z_2,u_2) = - 2 f_0^2 \Big(\frac{\xi_0}{\partial^2} (D_1^+)^4 \delta^{(2,-2)}(u_2,u_1)\nonumber\\
&&\qquad\qquad\qquad\qquad\qquad - \frac{\xi_0-1}{\partial^4} (D_1^+)^4 (D_2^+)^4 \frac{1}{(u_1^+ u_2^+)^2}\Big) \delta^6(x_1-x_2) \delta^8(\theta_1-\theta_2)\delta^{AB};\qquad\\
\label{Hypermultiplet_Propagator_Non_Abelian}
&& (G_q^{(1,1)})_i{}^j(z_1,u_1;z_2,u_2) = (D_1^+)^4 (D_2^+)^4 \frac{1}{\partial^2} \delta^{14}(z_1-z_2) \frac{1}{(u_1^+ u_2^+)^3}\delta_i{}^j\,.
\end{eqnarray}

\begin{figure}[h]
\begin{picture}(0,2)
\put(1,0.5){\includegraphics[scale=0.2]{propagator_v.eps}}
\put(0.5,1.3){(1)}
\put(5,0.5){\includegraphics[scale=0.2]{propagator_q.eps}}
\put(4.5,1.3){(2)}
\put(9,0.5){\includegraphics[scale=0.2]{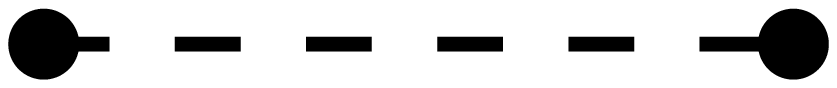}}
\put(8.5,1.3){(3)}
\put(13,0.5){\includegraphics[scale=0.2]{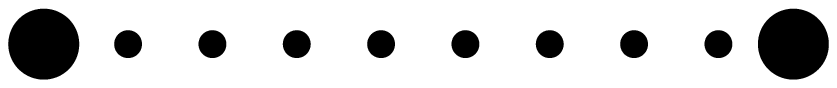}}
\put(12.5,1.3){(4)}
\end{picture}
\caption{The lines (1), (2), (3), and (4) denote the propagators of the gauge, hypermultiplet, Faddeev--Popov and Nielsen--Kallosh ghost superfields.}\label{Figure_Propagators_Non_Abelian}
\end{figure}

\noindent
In the explicit calculations in the non-abelian case we will use only the Feynman gauge $\xi_0=1$,
because under this choice the gauge propagator (\ref{Gauge_Propagator_Non_Abelian}) has the simplest form.
The propagators (\ref{Gauge_Propagator_Non_Abelian}) and (\ref{Hypermultiplet_Propagator_Non_Abelian}) will be graphically  denoted, as in the abelian case,
by the wavy and solid lines (see Fig. \ref{Figure_Propagators_Non_Abelian}). Also we will need the ghost propagators. They have the same form for
both the Faddeev--Popov and the Nielsen--Kallosh ghosts,

\begin{equation}
\frac{(D_1^+)^4 (D_2^+)^4}{2\partial^2}
\delta^{14}(z_1-z_2) \frac{(u_1^- u_2^-)}{(u_1^+ u_2^+)^3}\delta^{AB}
\end{equation}

\noindent
and will be depicted by the dashed and dotted lines, respectively.

\noindent
Finally, the propagator of the superfields $\xi^{(+4)}$ and $\sigma$ introduced in (\ref{Determinant_Remaining}) has the form

\begin{equation}
-\frac{(D_1^+)^4}{2\partial^2} \delta^{14}(z_1-z_2) \delta^{(0,0)}(u_1,u_2)\delta^{AB}.
\end{equation}

The interaction vertices can be easily read off from the interaction terms in the action. It is important that in the non-abelian case
on the external legs there can appear the background gauge superfield. Such legs will be denoted by the bold wavy lines.
Due to the linear background-quantum splitting (\ref{Splitting_Background_Quantum}) all vertices can contain both quantum and background wavy lines.
Precisely as in the ${\cal N}=(1,0)$ supersymmetric electrodynamics, in the non-abelian theory only the  triple vertex describing
the interaction of the hypermultiplet with the gauge superfield is present (the gauge superfield can be either background or quantum).

{}From the action (\ref{Action_Sypersymmetric_Yang_Mills}) we observe that there are infinitely many vertices with the number $n\ge 3$ of
the gauge superfield lines (and with no lines of any other superfields). Note that the gauge-fixing term (\ref{Gauge_Fixing_Term_Non-Abelian})
also contributes to these vertices (in this case the legs of the background gauge superfield come from the bridge).

Due to the presence of two super-background covariant derivatives in  the ghost action (\ref{Action_Faddeev_Popov_Ghosts}), there are triple and quartic vertices
containing two ghost lines. These vertices can have no more than one line of the quantum gauge superfield $v^{++}$ and no more than two lines of the background
gauge superfield $\bm{V}^{++}$.

The superfields $\varphi$, $\xi^{(+4)}$ and $\sigma$ interact  with the background gauge superfield only. For the superfield $\varphi$
only the triple and quartic vertices are possible, while the vertices involving $\xi^{(+4)}$ and $\sigma$ can also contain an arbitrary number
of the background gauge superfields coming from the superfield $\bm{V}^{--}$ concealed in the operator $\stackrel{\bm{\frown}}{\bm{\Box}}$.

\subsection{One-loop divergences in harmonic superspace}
\label{Subsection_One_Loop_Non_Abelian}

In order to calculate the divergent part of the one-loop effective action, we again start from calculating divergences of the lowest-order Green functions
and then  restore the full result by the reasoning based on the unbroken background gauge invariance. This can be done as follows.
According to \cite{Bossard:2015dva}, on shell the one-loop logarithmic divergences  have the structure

\begin{equation}\label{Divergences_General}
\Gamma^{(1)}_{\infty,\ln} = \int d\zeta^{(-4)}\,du\,\Big[c_1 (F^{++ A})^2 +
i c_2  F^{++ A} (\widetilde q^+)^i (T^A)_i{}^j (q^+)_j + c_3 \Big((\widetilde q^{+})^i (q^+)_i\Big)^2\Big],
\end{equation}

\noindent where $c_i$ with $i=1,2,3$ are real numerical coefficients and the regularization by dimensional reduction is assumed.
The coefficients $c_i$ can be obtained by calculating the divergences of the two-point function of the background gauge superfield ($c_1$) and of the three-point
gauge-hypermultiplet function $(c_2)$. The coefficient $c_3$ vanishes,

\begin{equation}
c_3=0,
\end{equation}

\noindent
because the corresponding four-point hypermultiplet Green function is finite. Actually, in the non-abelian case the degree of divergence for diagrams
without external ghost legs is also given by the expression (\ref{Divergence_Degree_QED}). In the case of $L=1$, $N_q=4$, $N_D=0$
we obtain $\omega=-2$, for which reason the one-loop four-point hypermultiplet Green function is given by the convergent integrals.

For calculating the coefficient $c_1$ in the expression (\ref{Divergences_General}) we consider the two-point Green function of the background gauge superfield.
In the one-loop order it is contributed to by the superdiagrams presented in Fig. \ref{Figure_Background_Gauge_Harmonic_Diagrams},
in which the external bold wavy lines correspond to the background gauge superfield. They were calculated in Ref. \cite{Buchbinder:2017ozh}.
The following result for the sum of the corresponding contribution to the effective action has been obtained there:

\begin{figure}[h]
\begin{picture}(0,4)

\put(0.5,3.7){$(1)$}
\put(0.5,2.3){\includegraphics[scale=0.4]{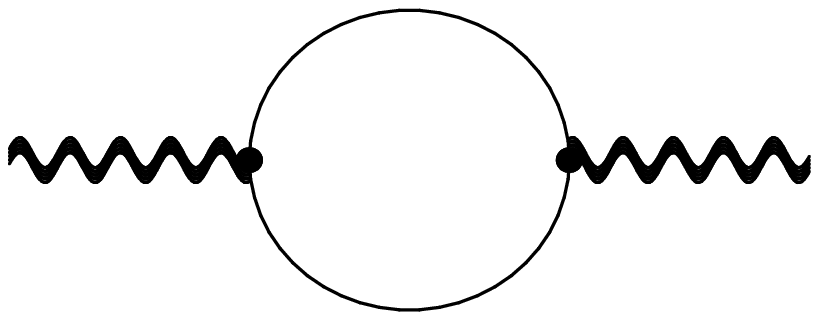}}
\put(4.5,3.7){$(2)$}
\put(4.5,2.3){\includegraphics[scale=0.4]{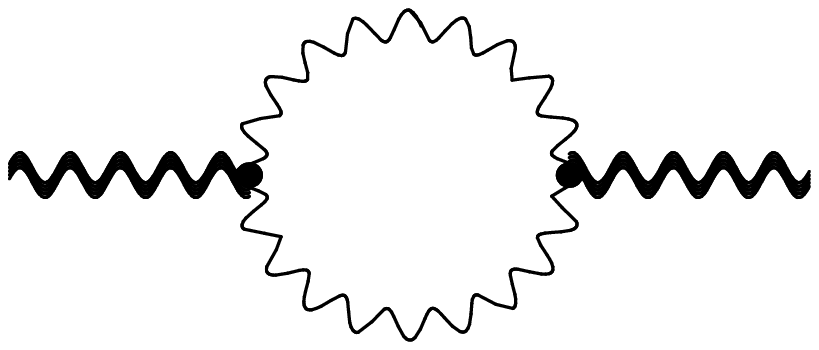}}
\put(8.5,3.7){$(3)$}
\put(8.5,2.3){\includegraphics[scale=0.4]{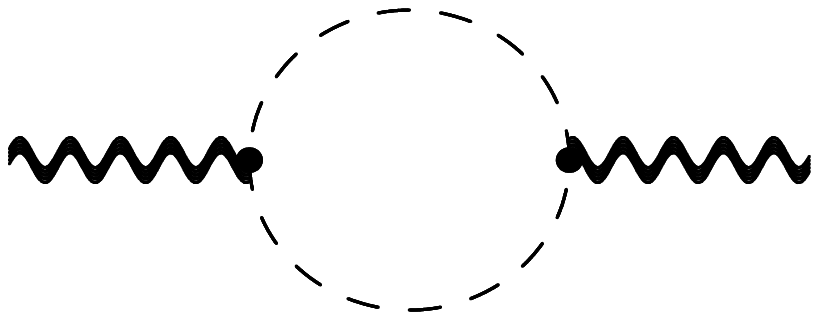}}
\put(12.5,3.7){$(4)$}
\put(12.5,2.3){\includegraphics[scale=0.4]{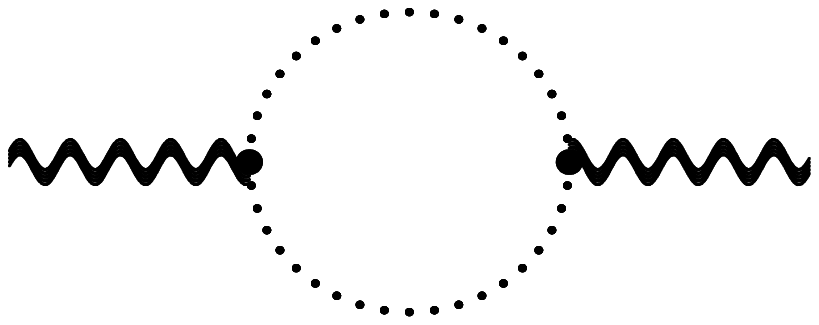}}

\put(2.8,1.5){$(5)$}
\put(3,0){\includegraphics[scale=0.4]{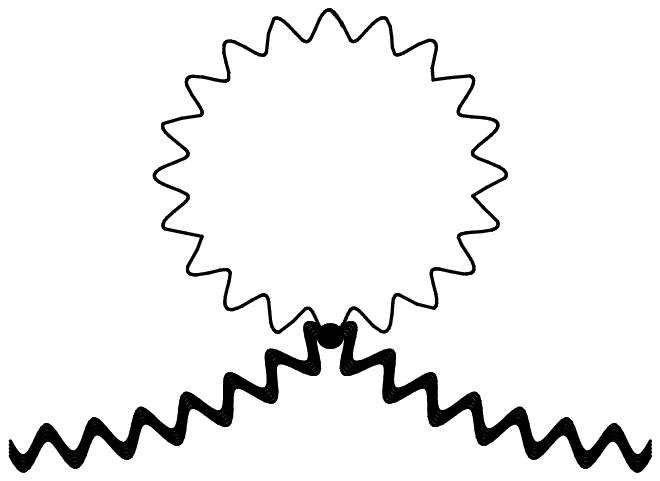}}
\put(6.8,1.5){$(6)$}
\put(7,0){\includegraphics[scale=0.4]{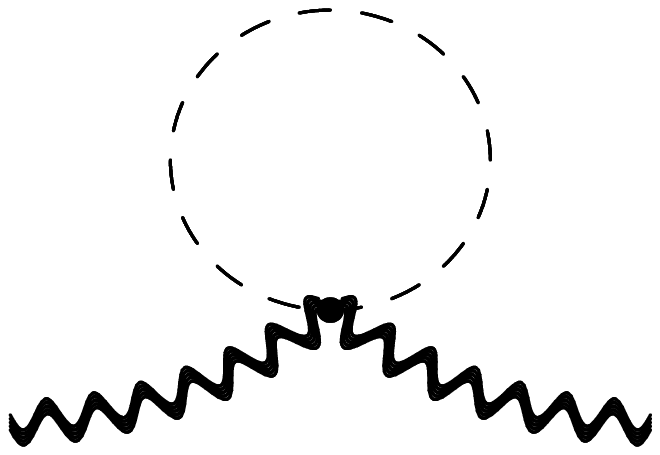}}
\put(10.8,1.5){$(7)$}
\put(11,0){\includegraphics[scale=0.4]{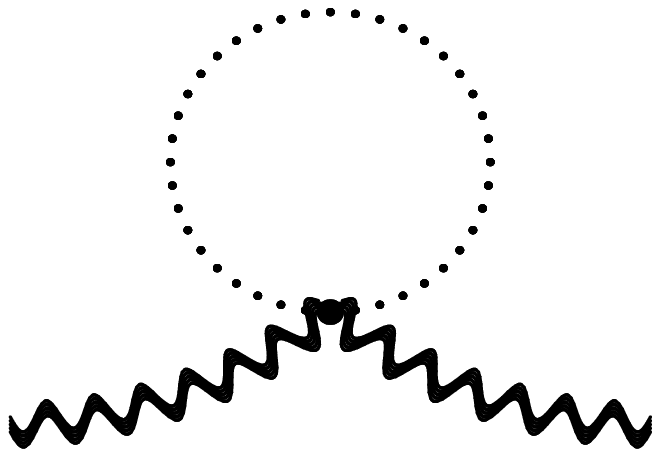}}

\end{picture}
\caption{Harmonic supergraphs representing the one-loop two-point Green
function of the background gauge superfield.}
\label{Figure_Background_Gauge_Harmonic_Diagrams}
\end{figure}

\begin{equation}\label{Two_Point_Background__Result}
\frac{i}{2} \Big[C_2 - T(R)\Big] \int \frac{d^6p}{(2\pi)^6} \int d^8\theta\,
du_1\, du_2\, \bm{V}^{++A}(p,\theta,u_1) \bm{V}^{++A}(-p,\theta,u_2)\frac{1}{(u_1^+ u_2^+)^2} \int \frac{d^6k}{(2\pi)^6} \frac{1}{k^2 (k+p)^2}.
\end{equation}

\noindent
This expression is divergent, the leading divergence being quadratic. However, the dimensional reduction can catch only the logarithmical divergences which can be written as

\begin{equation}\label{DRED_Background_Divergences_Non_Abelian}
\frac{1}{6\varepsilon (4\pi)^3} \Big[C_2 - T(R)\Big] \int d^{14}z\,du_1\, du_2\, \bm{V}^{++ A}(z,u_1) \partial^2 \bm{V}^{++ A}(z,u_2) \frac{1}{(u_1^+ u_2^+)^2}.
\end{equation}

\noindent
To calculate the quadratic divergences, one is led to use a regularization with an ultraviolet cut-off $\Lambda$. Then the leading quadratically divergent
terms are represented by the expression

\begin{equation}\label{Quadratic_Divergences_Non_Abelian}
-\frac{\Lambda^2}{4(4\pi)^3}\Big[C_2-T(R)\Big]\int d^{14}z\, du_1\, du_2\, \bm{V}^{++ A}(z,u_1) \bm{V}^{++ A}(z,u_2) \frac{1}{(u_1^+ u_2^+)^2},
\end{equation}

\noindent
while the logarithmical ones are obtained from (\ref{DRED_Background_Divergences_Non_Abelian}) via the substitution $1/\varepsilon \to \ln\Lambda$.

It is worth to note that the gauge invariant result in the non-abelian case also contains higher degrees of $\bm{V}^{++}$, which are encoded in (\ref{Divergences_General}).
Comparing the expression(\ref{DRED_Background_Divergences_Non_Abelian}) with

\begin{equation}
\int d\zeta^{(-4)}\,du\, (\bm{F}^{++ A})^2 = \int d^{14}z\, du_1\, du_2\, \frac{1}{(u_1^+ u_2^+)^2} \bm{V}^{++ A}(z,u_1) \partial^2 \bm{V}^{++ A}(z,u_2) + O\Big((\bm{V}^{++})^3\Big),
\end{equation}

\noindent
we obtain

\begin{equation}
c_1 = \frac{C_2-T(R)}{6\varepsilon (4\pi)^3},
\end{equation}

\noindent
which implies that, in the case of employing the dimensional reduction regularization, the divergent part of the one-loop effective action can be written as

\begin{equation}
\frac{C_2-T(R)}{3\varepsilon (4\pi)^3} \mbox{tr} \int d\zeta^{(-4)}\, du\, (\bm{F}^{++})^2 + \mbox{terms containing hypermultiplets}.
\end{equation}

\noindent
As for the quadratic divergences (\ref{Quadratic_Divergences_Non_Abelian}), they correspond to the lowest term in the power expansion of the gauge invariant object

\begin{equation}
-\Big[C_2-T(R)\Big]\frac{f_0^2 \Lambda^2}{(4\pi)^3}\, S_{\mbox{\scriptsize SYM}}[\bm{V}^{++}],
\end{equation}

\noindent where $S_{\mbox{\scriptsize SYM}}$ is given by
(\ref{Action_Sypersymmetric_Yang_Mills}).

The two-point Green function of the hypermultiplet is calculated similarly to the abelian case already
considered earlier. For non-abelian theories it is also determined by a single logarithmically divergent supergraph presented
in Fig. \ref{Figure_Hypermultiplet_2Point}. The only novelty is the presence of the hypermultiplet indices and the factor $C(R)_i{}^j$.
Exactly as in the abelian case, in the Feynman gauge $\xi_0=1\,$ the two-point Green function of the hypermultiplet vanishes (recall (\ref{Hypermultiplet_Green_Function})).

The coefficient $c_2$ in the expression (\ref{Divergences_General}) can be found by calculating the one-loop contribution to the three-point gauge-hypermultiplet
Green function, which is determined by two harmonic supergraphs presented in Fig. \ref{Figure_One-Loop_Vertex}.
The details of the calculation can be found in Ref. \cite{Buchbinder:2017ozh}, while here we provide only the answers:

\begin{figure}[h]
\begin{picture}(0,3)
\put(3.9,0.){\includegraphics[scale=0.18]{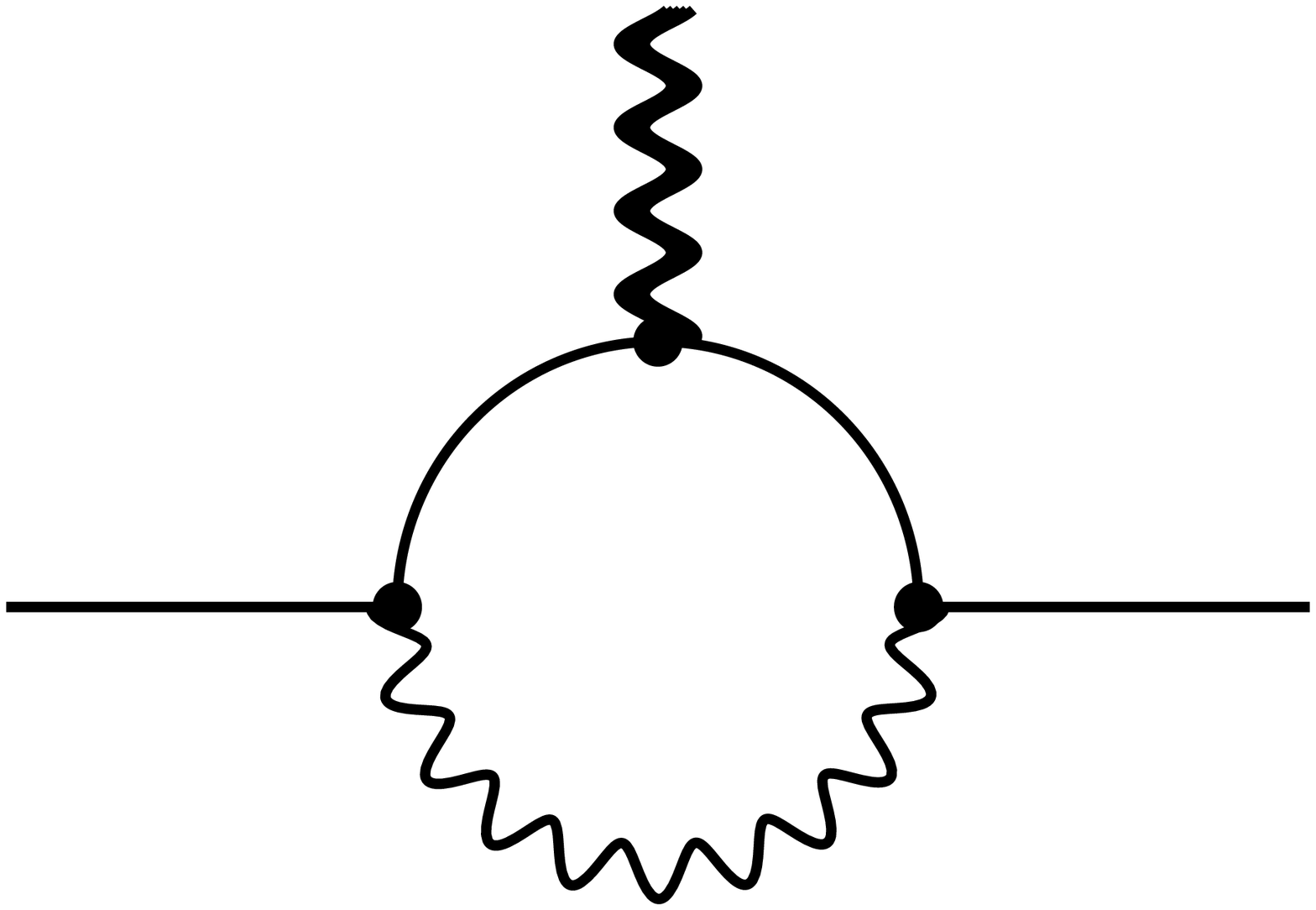}}
\put(3.9,2){$(1)$}
\put(8.9,0.15){\includegraphics[scale=0.18]{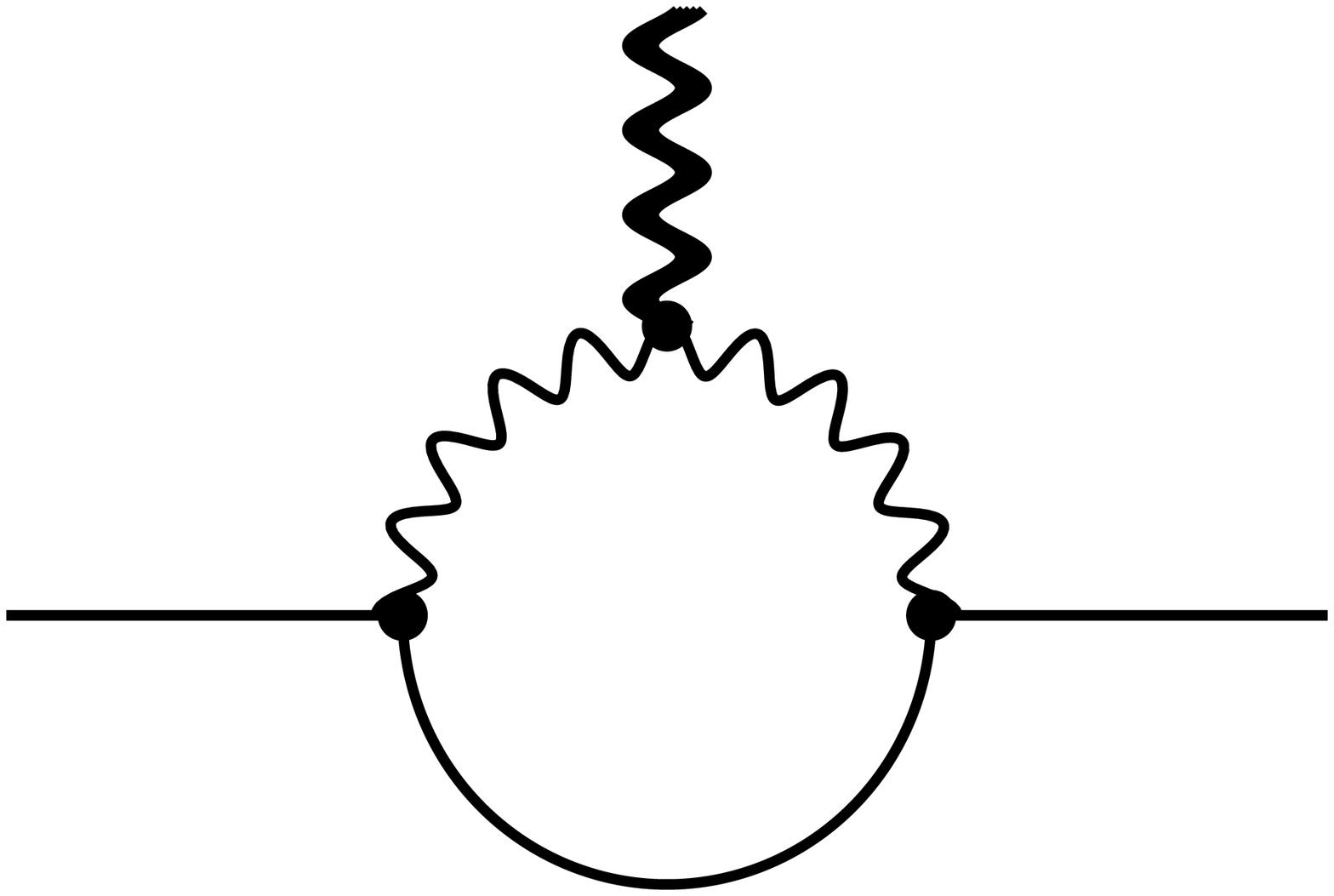}}
\put(8.9,2){$(2)$}
\end{picture}
\caption{These two harmonic supergraphs determine the three-point gauge-hypermultiplet function in the one-loop approximation.}\label{Figure_One-Loop_Vertex}
\end{figure}

\begin{eqnarray}
&&\hspace*{-5mm} (1) = -2f_0^2 \int \frac{d^{6}p}{(2\pi)^6}\,\frac{d^{6}q}{(2\pi)^6}\,d^8\theta\,du_1\,du_2\,
(\widetilde q^+)^i(q+p,\theta,u_1)
\Big[C(R)_i{}^k-\frac{1}{2} C_2\delta_i^k\Big] \bm{V}^{++}(-p,\theta,u_2)_k{}^j \nonumber\\
&&\hspace*{-5mm} \times (q^+)_j(-q,\theta,u_1)
\frac{1}{(u_1^+ u_2^+)^2} \int \frac{d^{6}k}{(2\pi)^6}\, \frac{1}{k^2 (q+k)^2 (q+k+p)^2};\\
&&\hspace*{-5mm} (2) =  f_0^2 C_2 \int \frac{d^6p}{(2\pi)^6}\, \frac{d^6q}{(2\pi)^6}\, d^8\theta\,du_1\,du_2\,
\widetilde q^+(q+p,\theta,u_1)^i \bm{V}^{++}(-p,\theta,u_2)_i{}^j q^+(-q,\theta,u_1)_j \nonumber\\
&&\hspace*{-5mm} \times  \frac{1}{(u_1^+ u_2^+)^2} \int \frac{d^6k}{(2\pi)^6}\, \frac{1}{k^2 (k+p)^2 (k+p+q)^2}\,.
\end{eqnarray}

\noindent
Obviously, both these expressions are logarithmically divergent. When using the regularization by dimensional reduction \cite{Siegel:1979wq}, the divergent part
of their sum is written as

\begin{equation}\label{Triple_Vertex_Divergence}
\frac{2if_0^2}{\varepsilon (4\pi)^3} \int \frac{d^{6}p}{(2\pi)^6}\,\frac{d^{6}q}{(2\pi)^6}\,d^8\theta\,du\,
(\widetilde q^+)^i(q+p,\theta,u) \Big[C(R)_i{}^k- C_2\delta_i^k\Big] \bm{V}_{\mbox{\scriptsize linear}}^{--}(-p,\theta,u)_k{}^j (q^+)_j(-q,\theta,u),
\end{equation}

\noindent
where

\begin{equation}
\bm{V}_{\mbox{\scriptsize linear}}^{--} \equiv \int du_1 \frac{\bm{V}^{++}(z,u_1)}{(u^+ u_1^+)^2}
\end{equation}

\noindent
is the lowest (linear) term in the expansion of $\bm{V^{--}}$ in powers of $\bm{V}^{++}$.

Rewriting the expression (\ref{Triple_Vertex_Divergence}) in the coordinate representation, we can cast it in the form

\begin{eqnarray}\label{Triple_Vertex_Divergence_Semicovariant}
&& \frac{2i f_0^2}{\varepsilon(4\pi)^3}\int d^{14}z\, du\, (\widetilde q^+)^i
\Big[C(R)_i{}^k- C_2\delta_i^k\Big] (\bm{V}_{\mbox{\scriptsize linear}}^{--})_k{}^j (q^+)_j\nonumber\\
&&\qquad\qquad\qquad\quad = \frac{2i f_0^2}{\varepsilon(4\pi)^3}\int d\zeta^{(-4)}\, du\, (\widetilde q^+)^i
\Big[C(R)_i{}^k- C_2\delta_i^k\Big] (\bm{F}_{\mbox{\scriptsize linear}}^{++})_k{}^j (q^+)_j,\qquad
\end{eqnarray}

\noindent
where the linear part of $\bm{F}^{++}$ is denoted by

\begin{equation}
\bm{F}_{\mbox{\scriptsize linear}}^{++} \equiv (D^{+})^4 \bm{V}_{\mbox{\scriptsize linear}}^{--}.
\end{equation}

\noindent
The expression (\ref{Triple_Vertex_Divergence_Semicovariant}) is the lowest term in the expansion of the gauge invariant expression

\begin{equation}\label{Triple_Vertex_Divergence_Covariant}
\frac{2i f_0^2}{\varepsilon(4\pi)^3}\int d\zeta^{(-4)}\, du\, (\widetilde q^+)^i
\Big[C(R)_i{}^k- C_2\delta_i^k\Big] \bm{F}^{++})_k{}^j (q^+)_j
\end{equation}

\noindent
in powers of $\bm{V}^{++}$. Comparing it with (\ref{Divergences_General}), we conclude that

\begin{equation}
c_2 = 2 f_0^2\, \frac{C(R)-C_2}{(4\pi)^3 \varepsilon}.
\end{equation}

Thus, when using the regularization by dimensional reduction, the total divergent part of the one-loop effective action for an arbitrary $6D$, ${\cal N}=(1,0)$ gauge theory
can be written as

\begin{eqnarray}\label{Total_Divergence_DRED}
&& (\Gamma^{(1)}_{\infty})_{\mbox{\scriptsize DRED}} = \frac{C_2-T(R)}{3\varepsilon (4\pi)^3} \mbox{tr} \int d\zeta^{(-4)}\, du\, (\bm{F}^{++})^2\nonumber\\
&&\qquad\qquad\qquad -\,  2i f_0^2 \frac{1}{\varepsilon(4\pi)^3}\int d\zeta^{(-4)}\, du\, \widetilde q^+ \big[C_2-C(R)\big]\bm{F}^{++} q^+.\qquad
\end{eqnarray}

\noindent
This is a final result for one-loop divergences. We see that in the
${\cal N}=(1,1)$  theory, where $T(Adj)=C_{2}$ and
$C(Adj)_{i}{}^{j}=C_{2}\delta_{i}{}^{j}$, the all one-loop
divergences are absent off-shell. This result was obtained in the
framework of the symersymmetric dimensional regularization.

However, it is interesting to understand how such a  result depends
on the regularization This is the reason why it is instructive to
study the one-loop divergences in the framework of some another
regularization. Here we present the corresponding result in the
regularization by an ultraviolet cut-off $\Lambda$. In this case it
is possible to calculate both quadratic and logarithmical one-loop
divergences,

\begin{eqnarray}\label{Gamma_CutOff}
&& (\Gamma^{(1)}_{\infty})_{\mbox{\scriptsize UV cut-off}} = -\big[C_2-T(R)\big]\frac{f_0^2 \Lambda^2}{(4\pi)^3}\, S_{SYM}[\bm{V}^{++}]
+ \ln\Lambda\Big[\frac{C_2-T(R)}{3 (4\pi)^3} \mbox{tr}\int d\zeta^{(-4)}\, du\, \qquad\nonumber\\
&& \times (\bm{F}^{++})^2 - 2i f_0^2 \frac{1}{(4\pi)^3}\int d\zeta^{(-4)}\, du\, \widetilde q^+ \big[C_2-C(R)\big] \bm{F}^{++} q^+\Big].
\end{eqnarray}

\noindent
Now we get the additional divergent term $S_{SYM}[\bm{V}^{++}]$ in
comparison with divergences within the dimensional regularization.
Nevertheless in the ${\cal N}=(1,1)$ theory this divergent terms also vanishes. Note that using of the cut-off
regularization can lead to some problems in higher loops. Actually, because of
a possible violation of the BRST invariance, the Slavnov--Taylor
identities \cite{Taylor:1971ff,Slavnov:1972fg} can be broken at the
quantum level (see, e.g., the calculation for supersymmetric
theories in Ref. \cite{Shakhmanov:2017wji}). However, these
identities can be restored with the help of a special subtraction
scheme, similar to the one constructed in
\cite{Slavnov:2002ir,Slavnov:2003cx}. Moreover, the BRST symmetry
guarantees the stability of the background-quantum splitting
(\ref{Splitting_Background_Quantum}). For non-invariant
regularizations this equation can receive some quantum corrections.
Nevertheless, in the one-loop approximation for the considered part
of the effective action all these problems are not essential. To
overcome them in higher loops, it is necessary to use an invariant
regularization, e.g., some versions of the higher covariant
derivative regularization \cite{Slavnov:1971aw,Slavnov:1972sq} in
the harmonic superspace (see \cite{Buchbinder:2015eva}).

As we already pointed out, with taking into account the relations
(\ref{Adjoint_Group_Factors}) we get that in $6D$, ${\cal N}=(1,1)$ SYM theory
all the divergences (including the quadratic ones) vanish.\footnote{The cancelation of quadratic divergences is also suggested by their relationship
with the (vanishing) divergences of $4D$, ${\cal N}=4$ theory.} In the gauge sector this occurs, because both quadratic and logarithmical divergences are proportional to $C_2-T(R)$. This result agrees with the calculation made earlier in \cite{Kazakov:2002nf,Kazakov:2002jd}, where the divergences in the gauge sector have been found using the component formulation of the theory. However, we also demonstrated that the divergences in the hypermultiplet sector vanish as well, if the theory is quantized in the manifestly ${\cal N}=(1,0)$ supersymmetric and gauge invariant way, and the Feynman gauge condition is used.

\subsection{Two-loop divergent part of the hypermultiplet two-point Green function of $6D$ SYM theories}
\label{Section_Two_Loop_hypermultiplet_Divergences}

\begin{figure}[h]
\begin{picture}(0,4.4)
\put(2.5,2.3){\includegraphics[scale=0.17]{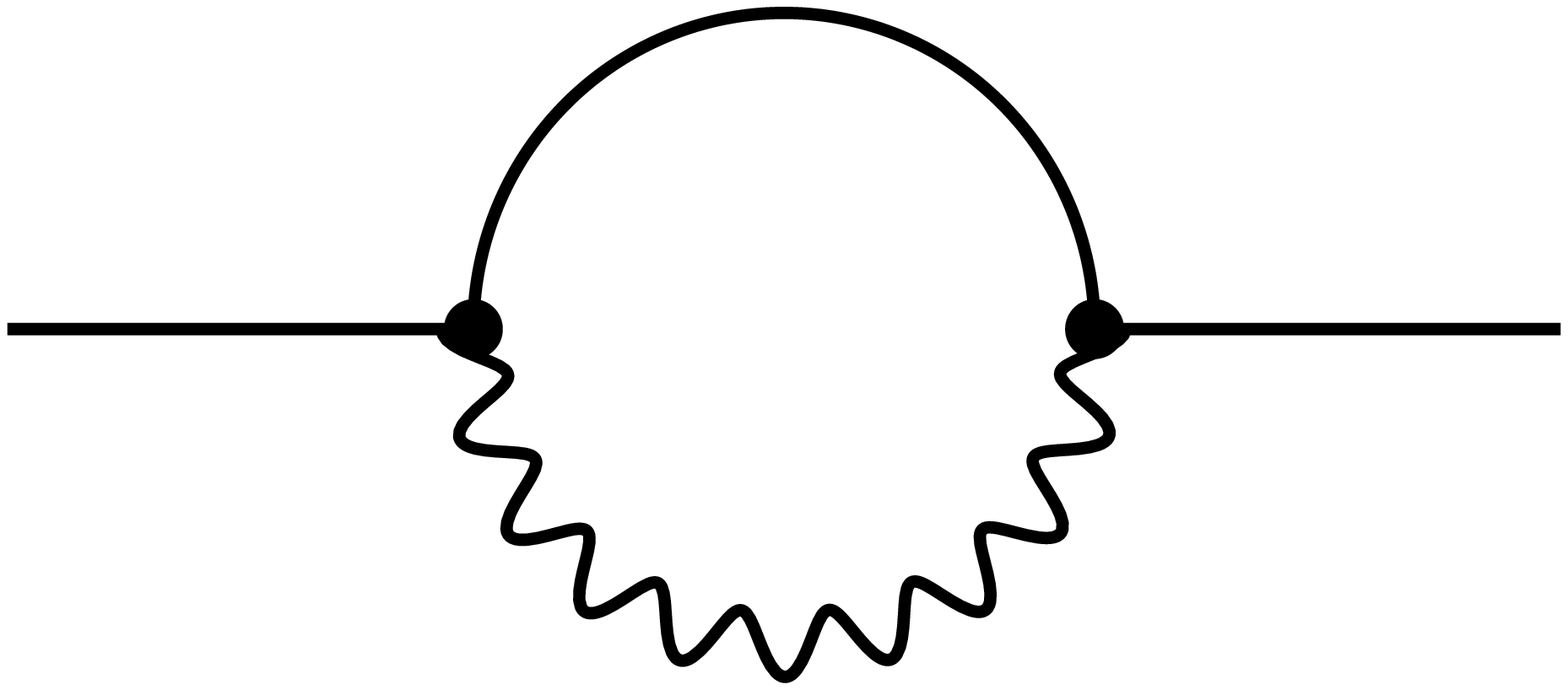}}
\put(6.5,2.3){\includegraphics[scale=0.17]{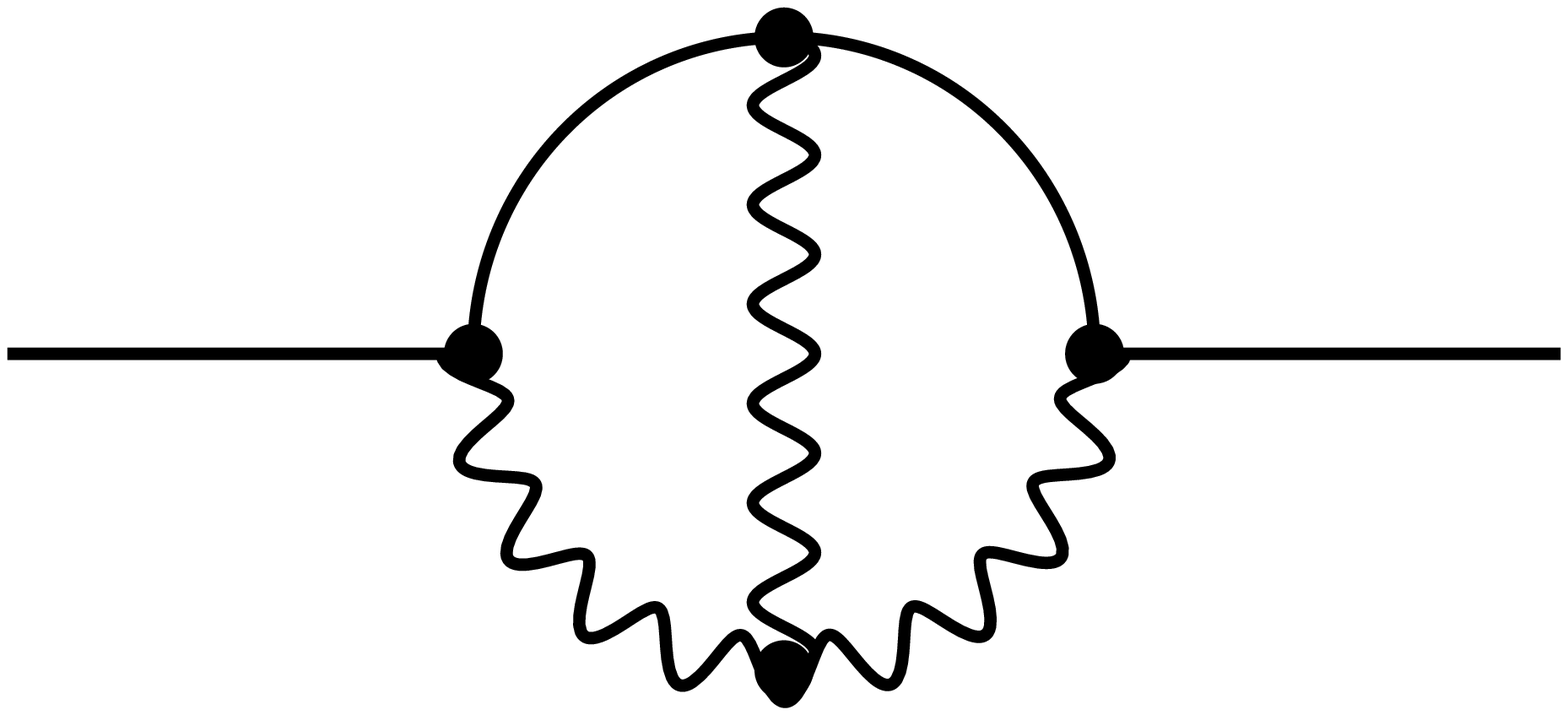}}
\put(10.5,2.3){\includegraphics[scale=0.17]{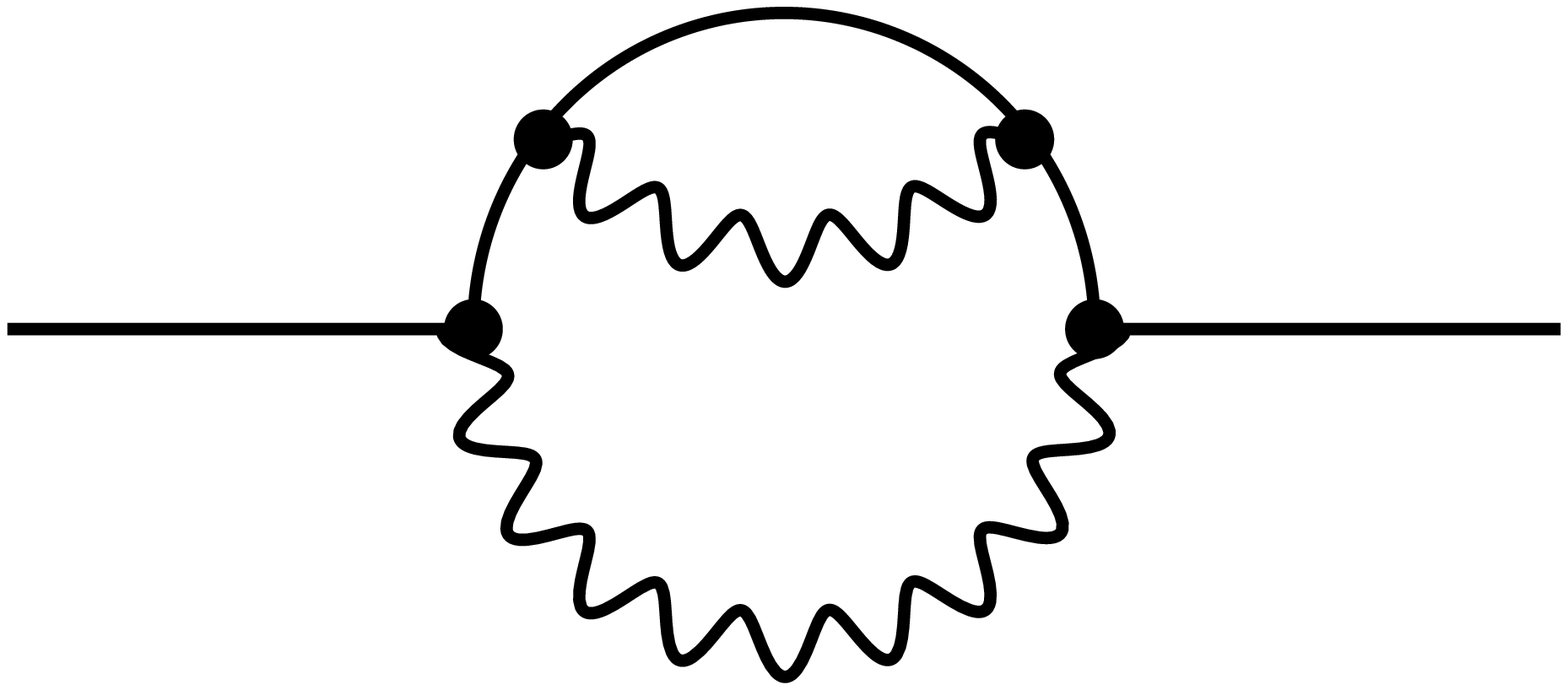}}
\put(4.5,0){\includegraphics[scale=0.17]{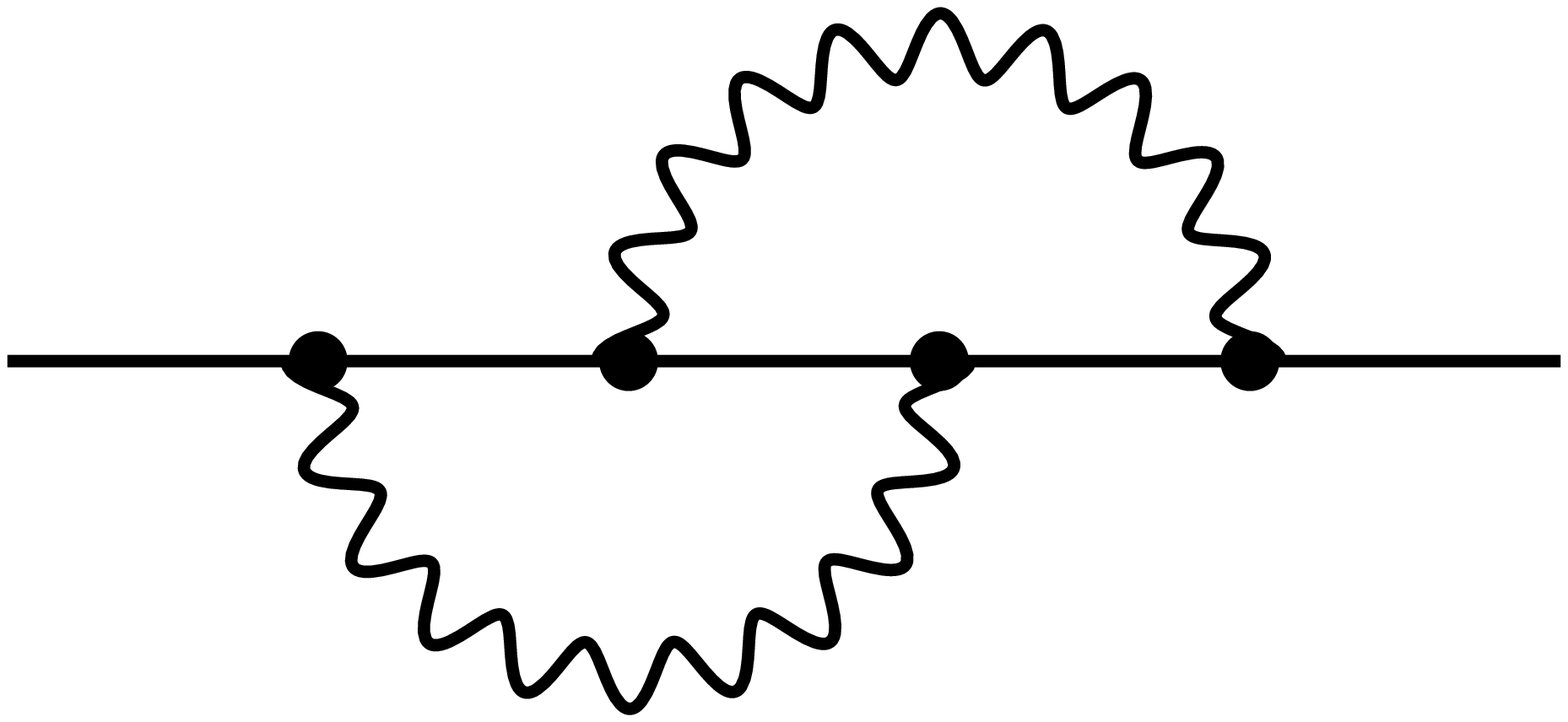}}
\put(8.5,-0.1){\includegraphics[scale=0.17]{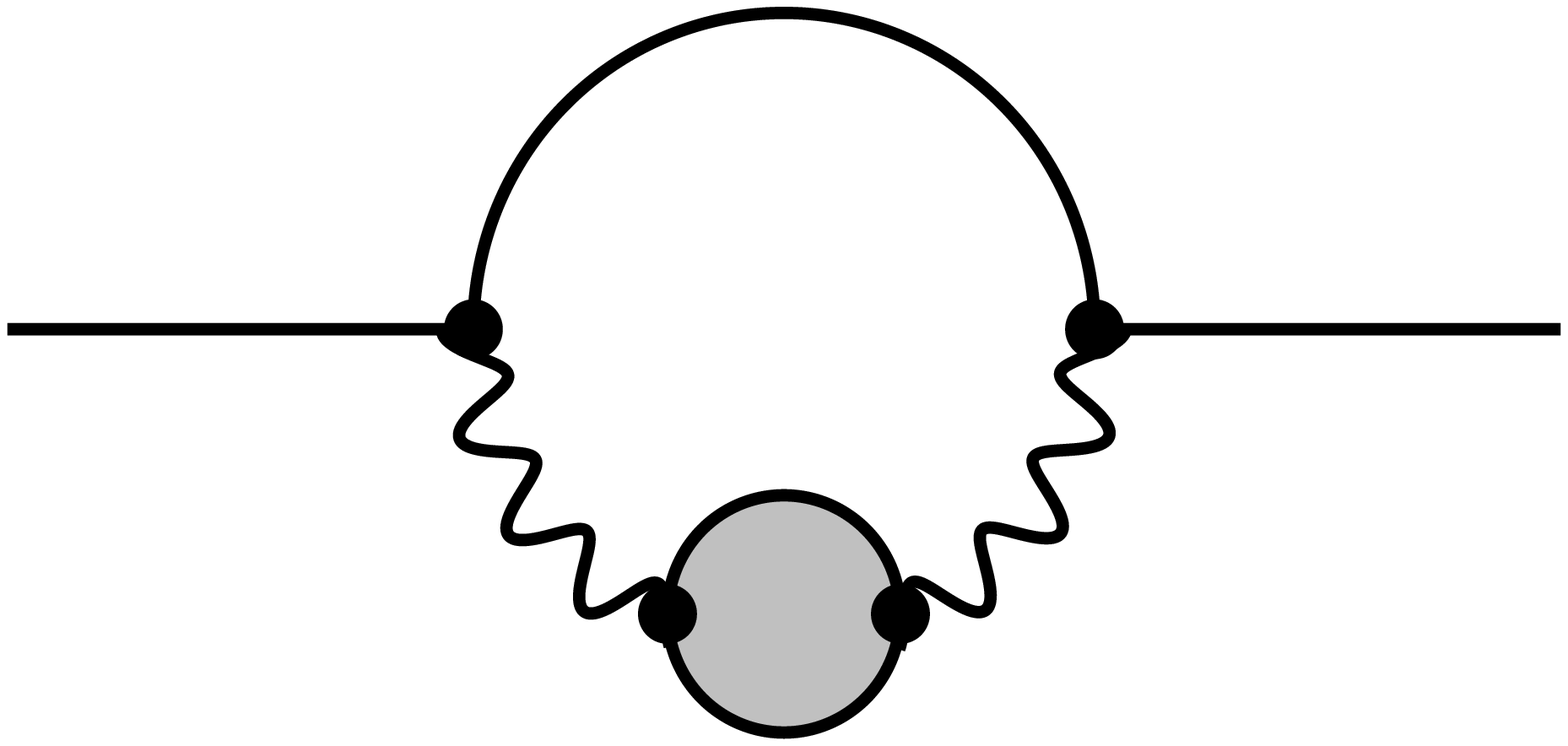}}
\put(2.4,3.7){$(1)$}
\put(6.4,3.7){$(2)$}
\put(10.5,3.7){$(3)$}
\put(4.5,1.3){$(4)$}
\put(8.5,1.3){$(5)$}
\end{picture}
\caption{Supergraphs representing the two-point hypermultiplet Green function in the two-loop approximation.}\label{Figure_2Loop_Diagrams}
\end{figure}

\begin{figure}[h]
\begin{picture}(0,4)
\put(2.2,2.0){\includegraphics[scale=0.17]{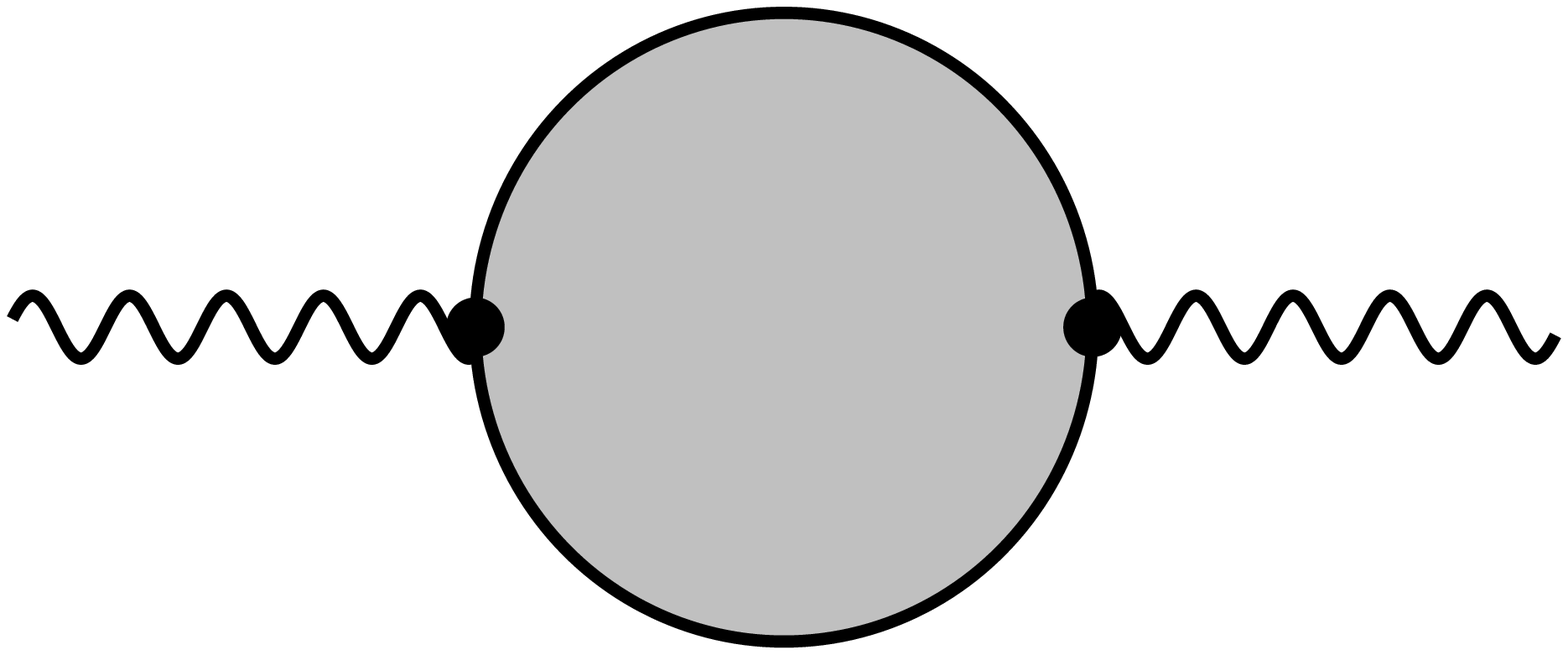}}
\put(5.9,2.6){$=$}
\put(10.5,2.0){\includegraphics[scale=0.4]{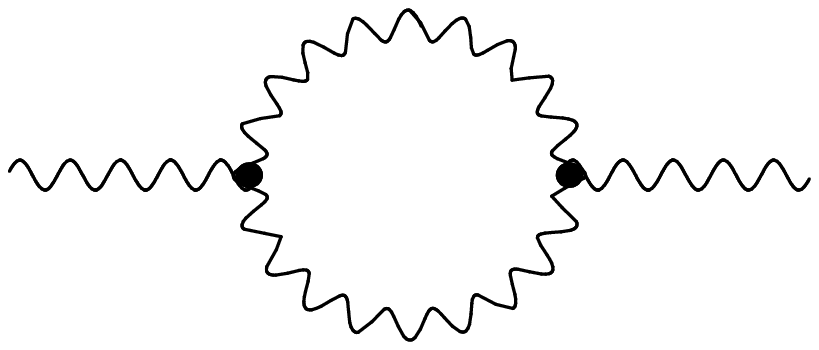}}
\put(9.6,2.6){$+$}
\put(6.7,1.7){\includegraphics[scale=0.4]{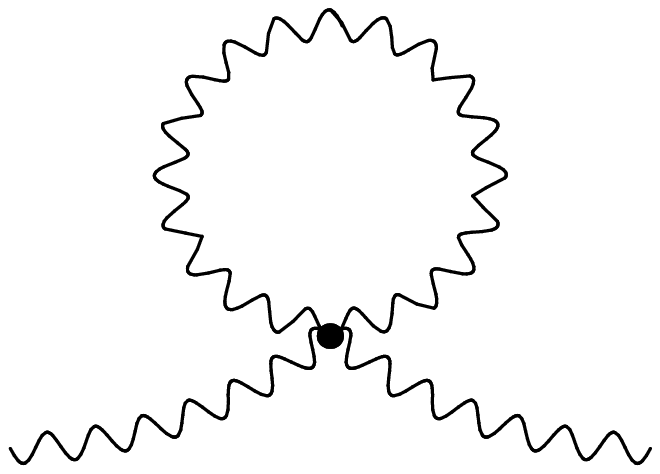}}
\put(6.2,-0.01){\includegraphics[scale=0.4]{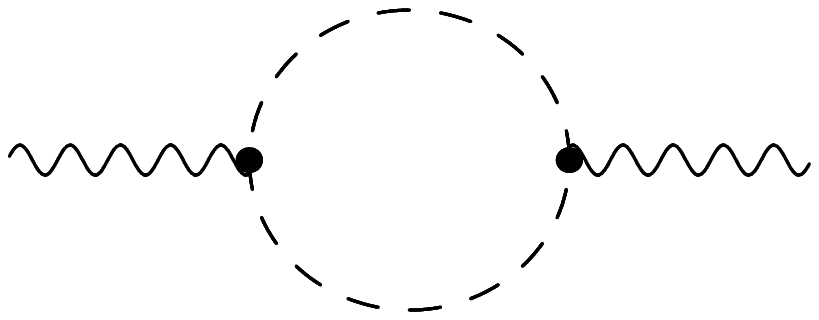}}
\put(10.6,-0.02){\includegraphics[scale=0.4]{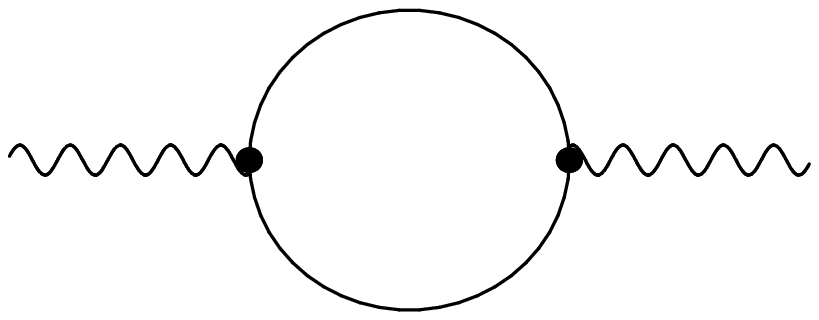}}
\put(5.5,0.5){$+$} \put(10.0,0.5){$+$}
\end{picture}
\caption{In Fig. \ref{Figure_2Loop_Diagrams} the gray circle corresponds to the one-loop polarization operator which is given by the sum of the
harmonic supergraphs depicted here.}\label{Figure_Effective_Diagram_Gray}
\end{figure}

The calculation of quantum corrections in the two-loop approximation is a much more complicated problem. To date,
the two-loop divergences in the harmonic superspace formalism have been found only for the two-point Green function
of the hypermultiplet. It is determined by the diagrams depicted in Fig. \ref{Figure_2Loop_Diagrams}.
In the diagram (5) in Fig. \ref{Figure_2Loop_Diagrams} the gray disk corresponds to the insertion of
the one-loop polarization operator of the quantum gauge superfield. It is given by the sum of the one-loop superdiagrams presented
in Fig. \ref{Figure_Effective_Diagram_Gray}. The details of the two-loop calculations can be found in Ref. \cite{Buchbinder:2017gbs}.
The formal result for the Green function under the consideration (without a regularization) is given by the expression (written in the Minkowski space before the Wick rotation)

\begin{eqnarray}\label{Two-Loop_Two-Point_Green_Function}
&& 4 f_0^4 \int \frac{d^6p}{(2\pi)^6} d^8\theta\, \int\frac{du_1\, du_2}{(u_1^+ u_2^+)}\, \Bigg[\widetilde q^+(p,\theta,u_1)^i \Big(-C(R)^2 + C_2 C(R)\Big)_i{}^j q^+(-p,\theta,u_2)_j\nonumber\\
&&\times \int \frac{d^6k}{(2\pi)^6} \frac{d^6l}{(2\pi)^6} \frac{1}{k^2 l^2 (k+l)^2 (k+l+p)^2 (k+p)^2}
+ \Big(C_2-T(R)\Big)\widetilde q^+(p,\theta,u_1)^i  C(R)_i{}^j\qquad\nonumber\\
&& \times q^+(-p,\theta,u_2)_j \int \frac{d^6k}{(2\pi)^6} \frac{d^6l}{(2\pi)^6} \frac{1}{k^4 (k+p)^2 l^2 (k+l)^2} \Bigg].
\end{eqnarray}

\noindent
In agreement with Eq. (\ref{Divergence_Degree_QED}) this Green function is quadratically divergent. The regularization by dimensional reduction cannot be used for calculating
the quadratic divergences, so it is necessary to use different regularization schemes. However, let us consider ${\cal N}=(1,1)$ SYM theory,
with the hypermultiplet in the adjoint representation, $R=Adj$. Using Eq. (\ref{Adjoint_Group_Factors}), we observe that
the expression (\ref{Two-Loop_Two-Point_Green_Function}) for this theory vanishes identically. This implies that the leading quadratic divergences
are canceled out and the total divergences can be calculated, based on the dimensional reduction. However, even after the replacement $6\to D$ the
expression (\ref{Two-Loop_Two-Point_Green_Function}) vanishes. Therefore, the considered Green function for ${\cal N}=(1,1)$ SYM theory vanishes identically.
Taking into account that ${\cal N}=(1,1)$ supersymmetry intertwines the gauge and hypermultiplet superfields, it is reasonable to suggest that
all two-point Green functions of this theory also vanish identically.

Nevertheless, two-loop off-shell divergences may arise in the four-point Green functions. To see this, it is sufficient to calculate the
four-point Green function of the hypermultiplet. This work is in progress now.

\subsection{Manifestly gauge covariant analysis}
\label{Manifestly gauge covariant analysis}

In this Section we briefly discuss how the proper-time method
can be used for analysis of divergent contributions in $6D$
$\cN=(1,0)$ SYM theory \eq{Action}. After splitting the
superfields $V^{++}, q^{+}$ into the sum of the background parts
$\bm{V}^{++}, \bm{Q}^{+}$ and the quantum parts $v^{++},
q^{+}\,$,
 \be
 V^{++}\to \bm{V}^{++} + v^{++}, \qquad q^{+} \to \bm{Q}^{+} + q^{+}\,,
 \ee
we expand the full action  in a power series in quantum superfields. In
the one-loop order, the first quantum correction to the
classical action, $\Gamma^{(1)}[\bm{V}^{++}, \bm{Q}^{+}]\,$, is given by the
following functional integral \cite{Buchbinder:2001wy,Buchbinder:2006td}:
 \be
 e^{ i\Gamma^{(1)}[\bm{V}^{++}, \bm{Q}^{+}]} =\mbox{Det}^{1/2}\sB \int {\cal
D}v^{++}\,{\cal D}q^+\, {\cal D}{\bf b}\,{\cal D}{\bf c}\,{\cal
D}\varphi\,\,\, e^{iS_{2}[v^{++}, q^+, {\bf b}, {\bf c}, \varphi,
\bm{V}^{++}, \bm{Q}^{+}]}\,.
 \label{Gamma0}
 \ee
In this expression, the full quadratic (with respect to the quantum superfields) action $S_2$ is the sum of
three terms, namely, the classical action \eq{Action} in which the
background-quantum splitting was performed, the gauge-fixing term
\eqref{Gauge_Fixing_Term_Non-Abelian} and the ghost actions \eqref{Action_Faddeev_Popov_Ghosts} and
\eqref{Action_Nielsen-Kallosh_Ghosts}. The action $S_2$ contains the
mixed term of quantum vector multiplet and hypermultiplet. After
diagonalization we obtain the following one-loop contribution to
the effective action
 \bea
 \Gamma[\bm{V}^{++}, \bm{Q}^{+}]&=&\frac{i}{2}\mbox{Tr}\ln\Big\{ \sB^{AB} -
 2f_0^2 \widetilde{\bm{Q}}^{+\,i} \big( T^A G_q^{(1,1)}T^B\big)_i{}^{j} \bm{Q}^{+}_{j}\Big\}
 -\frac{i}{2}\mbox{Tr}\ln\sB \nn \\
&& -i\mbox{Tr}\ln(\nb^{++})^2_{\rm Adj}
 +\frac{i}{2}\mbox{Tr}\ln(\nb^{++})^2_{\rm Adj}
 +i\mbox{Tr}\ln\nb^{++}_{\rm R}\,, \label{1loop_div}
 \eea
where $G_q^{(1,1)}(1|2)$ is the background-dependent hypermultiplet
Green function \eq{Hypermultiplet_Propagator_Non_Abelian}. Also we
introduce the covariant d'Alembertian $\sB= \f12 (D^+)^4 (\nb^{--})^2$.
On the analytic superfields $\sB$ is reduced
to
\begin{eqnarray}
&& \sB= \eta^{MN} \bm{\nabla}_M \bm{\nabla}_N +  \bm{W}^{+a} \bm{\nabla}^{-}_a + \bm{F}^{++}
\bm{\nabla}^{--} - \frac{1}{2}(\bm{\nabla}^{--} \bm{F}^{++})\,,
\label{Box_First_Part}
\end{eqnarray}
where $\eta^{MN} = {\rm diag} (1, -1, -1, -1, -1, -1)$ denotes the
six-dimensional Minkowski metric and $\bm{\nabla}_{M}=\pr_M+i\bm{A}_M$ is the
vector supercovariant derivative.

The $(\bm{F}^{++})^2$ part of the effective action depends only on the
background gauge superfield $\bm{V}^{++}$ and is given by the last
three terms in Eq. \eq{1loop_div}. More precisely,
 \bea
\Gamma_{\bm{F}^2}^{(1)}[\bm{V}^{++}] &=&-i\mbox{Tr}\ln(\bm{\nb}^{++})^2_{\rm Adj}
 +\frac{i}{2}\mbox{Tr}\ln(\bm{\nb}^{++})^2_{\rm Adj}
 +i\mbox{Tr}\ln\bm{\nb}^{++}_{\rm R} \nonumber\\
&=&-i\mbox{Tr}\ln\bm{\nb}^{++}_{\rm Adj}+i\mbox{Tr}\ln\bm{\nb}^{++}_{\rm R}\,.
 \label{1loopV}
 \eea
Here the index 'R' refers to the representation of hypermultiplet. Keeping in mind the explicit expressions for the covariant harmonic
derivatives, $(\bm{\nb}^{++}_{\rm R})^i_j= D^{++} \d^{i}_j +i (\bm{V}^{++})^C
(T^C)_i{}^{j}$ and $(\bm{\nb}^{++}_{\rm Adj})^{AB}= D^{++} \d^{AB}
-f^{ACB} (\bm{V}^{++})^C \,$, we vary the expression \eq{1loopV} with
respect to the background gauge superfield $(\bm{V}^{++})^A\,$
 \bea
 \d \Gamma_{\bm{F}^2}^{(1)}[\bm{V}^{++}] = i \Tr\, f^{ACB}\, \d
 (\bm{V}^{++})^C\, (G^{(1,1)})^{BA} - \Tr\,  (T^C)_j{}^{i}\, \d (\bm{V}^{++})^C
 \,(G^{(1,1)}_q)_i{}^j\,.
 \eea
Here $(G^{(1,1)}_{q})_i{}^j$ is the superfield Green function
\eq{Hypermultiplet_Propagator_Non_Abelian} for the operator
$(\bm{\nb}^{++})_i{}^{j}$ acting on the superfields in the
representation $R$ to which the hypermultiplet
belongs. Also we denoted  the Green function for
the operator $(\bm{\nb}^{++})^{BA}$, which acts on superfields in
adjoint representation, by $(G^{(1,1)})^{BA}$ . The Green function $(G^{(1,1)})^{BA}$ has
the structure similar to \eq{Hypermultiplet_Propagator_Non_Abelian}.

The background-dependent Green function $G^{(1,1)}_{q}(1|2)$
\eq{Hypermultiplet_Propagator_Non_Abelian} can be written as the
following proper-time integral
 \bea
G^{(1,1)}_{q}(1|2) = -\int_0^\infty
 d(is)(is\mu^2)^{\f\varepsilon2} e^{is \sB} (D^+_1)^4(D^+_2)^4
 \f{\d^{14} (z_1-z_2)}{(u_1^+u_2^+)^3}\,.
 \eea
Here $s$ is the proper-time parameter and $\mu$ denotes an arbitrary
regularization parameter with the dimension of mass. Our aim is to calculate
the divergent part of the effective action \eq{1loopV}. In the
proper-time regularization scheme (see, e.g.,
\cite{Buchbinder:1998qv}) the divergences correspond to the
pole terms of the form $1/\varepsilon\,$, $\varepsilon \to 0$,
with $D = 6 -\varepsilon$. Then, calculating the divergences according to the standard technique, after some
(rather non-trivial) transformations we obtain
 \be
 \Gamma^{(1)}_{\bm{F}^2} =  \f{C_2 - T(R)}{6(4\pi)^3 \varepsilon}\,
 \int d\zeta^{(-4)} du\, (\bm{F}^{++ A})^2\, =  \f{C_2 - T(R)}{3(4\pi)^3 \varepsilon}\,
 \mbox{tr}\int d\zeta^{(-4)} du\, (\bm{F}^{++})^2\,, \label{div7}
 \ee
where $\bm{F}^{++} = \bm{F}^{++ A} t^A$, with $t^A$ being
the fundamental representation generators.

The hypermultiplet-dependent part $\widetilde{\bm{Q}}^+
\bm{F}^{++}\bm{Q}^+$ of the one-loop counterterm comes from the
first term in \eq{1loop_div}. In order to find this contribution,
firstly we rewrite it as a sum of two terms,
 \bea
 &&\frac{i}{2}\mbox{Tr}\ln\Big\{ \sB^{AB} - 2f_0^2
 \widetilde{\bm{Q}}^{+\,i} \big( T^A G^{(1,1)}_qT^B\big)_i{}^{j} \bm{Q}^{+}_{j}\Big\}
  = \frac{i}{2}\mbox{Tr}\ln \sB \nn \\
 &&\qquad\qquad \qquad +  \frac{i}{2}\mbox{Tr}\ln\Big\{ \d^{AB} - 2f_0^2
 (\sB^{-1})^{AC}\widetilde{\bm{Q}}^{+\,i}
 \big( T^C G^{(1,1)}_q T^B\big)_i{}^{j} \bm{Q}^{+}_{j}\Big\}\,. \label{QFQ1}
 \eea
Then, following \cite{Buchbinder:2016gmc}, we decompose the second
logarithm up to the first order and compute the functional trace
 \bea
 \Gamma^{(1)}_{\bm{QFQ}} &=&  -i f_0^2\int d\zeta^{(-4)} du\,
 \widetilde{\bm{Q}}^{+\,j}\bm{Q}^{+}_{i}
 \, (\sB^{-1})^{AB } \big( T^B G^{(1,1)_q}T^A\big)_j{}^{i}\Big|^{2=1}_{\rm  div} \nonumber\\
 &=& -i f_0^2\int d\zeta^{(-4)} du\,
 \widetilde{\bm{Q}}^{+\,i}\bm{Q}^{+}_{j} \\
&&\quad \times \, (\sB^{-1})^{AB} \big( T^B \sB^{-1}
T^A\big)_i{}^{j}
 (u^+_1u^+_2)\, \delta^{6}(x_1-x_2)\Big|_{2=1}\,.\nonumber
  \eea
Here we use of the explicit form of the Green function
$(G^{(1,1)}_q)_i{}^j$ \eq{Hypermultiplet_Propagator_Non_Abelian} for
extracting the divergent contribution to the effective action. After this we
decompose the inverse $\sB^{-1}$ of the covariant operator $\sB$ \eq{Box_First_Part}
up to the second order and obtain
 \begin{equation}
\Gamma^{(1)}_{\bm{QFQ}}[\bm{V}^{++}, \bm{Q}^+] = -\f{ 2if_0^2}{(4\pi)^3 \varepsilon}
\int d\zeta^{(-4)}du\,
 \widetilde{\bm{Q}}^{+\,i}(C_2\delta_i^{l}-C(R)_i{}^{l})(\bm{F}^{++})^A\,\,
 (T^A)_l{}^{j}\,\bm{Q}^{+}_{j}. \label{div8}
 \end{equation}

Summing up the contributions \eq{div7} and \eq{div8}, we obtain the final result for the total divergent contribution
 \begin{eqnarray}
 &&\Gamma^{(1)}_{div}[\bm{V}^{++}, \bm{Q}^+] = \f{C_{2} - T(R)}{ 3(4\pi)^3 \varepsilon}\, \mbox{tr}\int d\zeta^{(-4)}du\,
 (\bm{F}^{++})^2\nonumber\\
 &&\qquad\qquad\qquad\quad - \frac{2if_0^2}{(4\pi)^3\varepsilon} \int d\zeta^{(-4)} du\, \widetilde{\bm{Q}}^+ (C_2-C(R)) \bm{F}^{++} \bm{Q}^+.
 \label{answer}
 \end{eqnarray}
We see that the result \eq{answer} derived by the manifestly gauge invariant
method, coincides with the previous result \eq{Total_Divergence_DRED} based on supergraph
calculations.

\subsection{Low-energy effective action}
\label{Subsection_Finite_Contributions}

The background field method developed in the previous sections is a
powerful tool for calculation of the finite contributions to
the effective action in a manifestly gauge invariant way.\footnote{For
background field method in $4D$ harmonic superspace  and  its
application to the problem of effective action in $\cN=2,4 $ SYM
theories see  papers
\cite{Buchbinder:1998np,Buchbinder:2001xy,Buchbinder:2002tb,Buchbinder:2001ui,Buchbinder:1999jn,Buchbinder:2016xeq}
and references therein.}. In this section we evaluate the finite
one-loop leading low-energy contribution to the effective action of
$6D,\,\cN=(1,1)$ SYM theory in the $\cN=(1,0)$
harmonic superspace formulation. An important aspect of the
consideration is the use of omega-hypermultiplet.

First, we formulate $6D, \,\cN=(1,1)$ SYM theory in
terms of $6D,\, \cN=(1,0)$ analytic harmonic superfields $V^{++}$
and $\omega$, which are the gauge supermultiplet and
the hypermultiplet, respectively. The action of $\cN = (1,1)$
SYM theory in this case reads
 \bea
S[V^{++}, q^+]&=& \frac{1}{f_0^2}\Big\{\sum\limits^{\infty}_{n=2}
\frac{(-i)^{n}}{n} \tr \int d^{14}z\, du_1\ldots du_n
\frac{V^{++}(z,u_1 ) \ldots
V^{++}(z,u_n ) }{(u^+_1 u^+_2)\ldots (u^+_n u^+_1 )}  \nn \\
&& - \f12 \tr \int d\zeta^{(-4)}\, \nb^{++}\omega \nb^{++}\omega
\Big\}\,,
\label{S0}
\eea
where
\bea
\nabla^{++}\omega = D^{++} \omega + i [V^{++},\omega]\,.
\nonumber
\eea
Here both $V^{++}$ and $\omega$ take the values in
the adjoint representation. The action \eq{S0} is
invariant under the infinitesimal gauge transformations
 \bea
 \d V^{++} = -\nb^{++} \Lambda\,,  \quad \d \omega =  i[\Lambda, \omega]\,,
 \label{gtr}
 \eea
where $\Lambda(\zeta, u) = \widetilde{\Lambda}(\zeta, u)$ is an analytic real
gauge parameter.

The action \eq{S0} was written in terms of $\cN=(1,0)$ harmonic
superfields. However, this action possesses an additional hidden $\cN=(0,1)$
supersymmetry realized by the transformations
 \bea
\delta V^{++} &=&  2 (\epsilon^{+ A}u^+_A) \omega - \nabla^{++}
\big((\epsilon^{+ A}u^-_A)
 \omega\big),\label{V++HidOm} \\
\delta \omega &=& i(\epsilon^{- A}u^-_A) F^{++} -i (\epsilon^A_a
u^-_A)\, W^{+ a} \label{omegaHid},
 \eea
where $A=1,2$ is the  Pauli-G\"{u}rsey $SU(2)$ index and
$W^{+a}=-\f{i}{6}\varepsilon^{abcd}D^+_b D^+_c D^+_d V^{--}\,,
D^{+}_a W^{+ a} = 4 F^{++}$. As a result, this action describes ${\cal N}=(1,1)$ SYM theory.

Our further consideration is based on the background field method in
six-dimensional $\cN=(1,0)$  harmonic superspace which was developed
in the previous subsection.  Here we focus only on aspects related
to omega-hypermultiplet. As in the previous sections, we represent
the original superfields $V^{++}$ and $\omega$ as a sum of the
background superfields $\bm{V}^{++}, \bm{\Omega}$ and the
quantum superfields $v^{++}, \omega\,$. In the present case it is
convenient to append the coupling constant $f_0$ in front of quantum
fields
 \be
 V^{++}\to \bm{V^{++}} + f_0 v^{++}, \qquad \omega
 \to \bm{\Omega} + f_0 \omega\,.
 \ee
Then we expand the action in a powers of the
quantum fields. The one-loop contribution to the effective action
$\Gamma^{(1)}$ for the model \eq{S0} is defined  by the quadratic part
of quantum action $S_2$,
  \bea
 S_2 &=& S_{\mbox{\scriptsize gh}}+ \frac{1}{2}\tr\int d\zeta^{(-4)}\, v^{++}\sB v^{++}
  -\f12 \tr\int d\zeta^{(-4)}\,  (\nb^{++} \omega)^2 \nn \\
 && - i\tr \int d\zeta^{(-4)}\Big\{
  \nb^{++}\omega [v^{++},\bm{\Omega}]
  + \nb^{++}\bm{\Omega} [v^{++},\omega] +\f{i}2[v^{++},\bm{\Omega}]^2\Big\} \,. \label{S2}
 \eea
The action $S_{\mbox{\scriptsize gh}}$ in \eq{S2} is a sum of the  action for
Faddeev-Popov ghosts ${\bf b}$ and ${\bf c}$
\eq{Action_Faddeev_Popov_Ghosts} and the action for Nielsen-Kallosh
ghost $\varphi$ \eq{Action_Nielsen-Kallosh_Ghosts}. The
covariantly-analytic operator $\sB$ \eq{Box_First_Part} depends on
the background gauge superfield.

The action \eq{S2} includes the background superfields $ \bm{V}^{++} $
and $ \bm{\Omega}$ which belong to the Lie algebra of gauge group. Let
us suppose that the gauge group of the theory \eq{S0} is $SU(N)$. For simplicity, we will also assume that the
background fields $\bm{V}^{++}$ and $\bm{\Omega}$ align in a fixed direction
in the Cartan subalgebra of $su(N)$
 \bea
 \bm{V}^{++} = V^{++}(\zeta,u) H\,, \qquad \bm{\Omega} = \Omega(\zeta,u)\,
 H\,. \label{bkg}
 \eea
Here $H$ is a fixed generator of  the Cartan subalgebra corresponding to some abelian subgroup $U(1)$. For our choice of the background superfields
the symmetry group of classical action
$SU(N)$ is broken down to $ SU(N-1)\otimes U(1)$. It is worth
to note that the pair of the background abelian superfields
$(V^{++},\Omega)$ forms the  abelian gauge $\cN=(1,1)$
multiplet. In the bosonic sector it contains a single real gauge
vector field $A_M(x)$ and four real scalar fields $\phi(x)$ and
$\phi^{(ij)}(x)\,, i,j=1,2$. The fields $\phi$ and
$\phi^{(ij)}$ are scalar components of the hypermultiplet $\Omega$
\cite{Galperin:2001uw}. It is known that the abelian vector field and four
scalars describe the bosonic world-volume degrees of freedom of a
single D5-brane in in six-dimensional space-time \cite{Giveon:1998sr,Blumenhagen:2006ci}.

According to the definition \eq{bkg}, the classical motion equations for the
background superfields $V^{++}$ and  $\Omega$ are reduced to the free
ones
 \bea
 F^{++} = 0\,, \qquad (D^{++})^2 \Omega = 0\,. \label{EqmB}
 \eea

In our further consideration we assume that the background
superfields satisfy  the classical equation of motion \eq{EqmB} and
also are slowly varying in space-time
 \bea
 \partial_M W^{+a} = 0\,, \qquad \partial_M \Omega = 0\,.
 \label{constBG}
 \eea

Since we assume that the background vector multiplet solves the
free equation of motion, $F^{++}=0$, the gauge superfield strength
$W^{+a}$ becomes an analytic superfield on shell. In the general case of
unconstrained background, $F^{++}\neq0$,  the superfield $W^{+a}$ is
non-analytic.

The transformations of the hidden $\cN=(0,1)$ supersymmetry for the gauge
superfield strength $W^{+a}$ and $\Omega$ \eq{omegaHid}, in
accordance with the conditions \eq{EqmB} and \eq{constBG}, have the
simple form
 \bea
  \delta \Omega = -i (\epsilon^A_a u^-_A)\, W^{+ a}\, \qquad \delta W^{+ a}=0.
  \label{Onshell2}
 \eea
Using \eq{Onshell2}, one can try to investigate the simplest $\cN=(1,1)$
invariants which can be obtained from the abelian analytic
superfields $W^{+a}$ and $\Omega$ under the assumptions \eq{EqmB}
and \eq{constBG}. It is easy to check that the following
gauge-invariant action,
 \bea
I= f_0^2 \int d\zeta^{(-4)} (W^{+})^4 {\cal F}(f_0 \Omega),
\label{I}
 \eea
is invariant under the transformation \eq{Onshell2}. Here we
introduced the fourth power of gauge superfield strength $(W^+)^4 =
-\f{1}{24}\varepsilon_{abcd} W^{+a}W^{+b}W^{+c}W^{+d}$. The function
${\cal F}(f_0\Omega)$ can in principle be arbitrary.  The simplest
choice, when the coupling constant $f_0$ is absent in the invariant, is ${\cal F} = \f{1}{{f_0^2}\Omega^2}$ in \eq{I}, which yields
 \bea
I_1= c\int d\zeta^{(-4)}\, \f{(W^{+})^4}{ \Omega^2}\,. \label{I1}
 \eea
The numerical coefficient $c$ in \eq{I1} cannot be fixed only by the symmetry
considerations and should be calculated using the quantum field theory methods.

So our next step is to find the constant $c$ by calculating the leading
low-energy contribution to the effective action of the theory
\eq{S0}. To perform the calculation we choose the Cartan-Weyl basis
for the $SU(N)$ generators. In this basis the quantum
superfield $v^{++}$  is decomposed as
 \bea
 v^{++} =  v^{++}_{\rm i} H_{\rm i}+ v^{++}_\a E_\a\,, \qquad {\rm i} = 1,..,
 N-1,\quad \a = 1,..,N(N-1)\,.
 \eea
For the generators  $E_\a$  corresponding to the root $\a$ we use
the normalization  $\tr(E_\a E_{-\b} )=
\d_{\a\b}$. The  Cartan subalgebra generators $H_{\rm i}$ satisfy
the relations $[H_{\rm i}, E_\a] = \a_{H_i} E_\a$. The integration
over quantum superfields $v^{++}$ and $\omega$ in \eq{Gamma0}
produces the one-loop effective action for the background
superfields $V^{++}$ and $\Omega$,
 \bea
\Gamma^{(1)}[V^{++},\Omega]&=& \f{i}2\Tr_{(2,2)} \ln\Big(\sB_H -
\alpha_H^2 \Omega^2\Big) + \f{i}2 \Tr \ln\Big[(\nb_H^{++})^2 +
A_{(+)}\f{\alpha^2_H}{\sB_H - \alpha_H^2 \Omega^2} A_{(-)}\Big]   \nn \\
&& - \f{i}2\Tr_{(4,0)} \ln \sB_H -i\Tr\ln (\nb_H^{++})^2 + \f{i}2
\Tr\ln (\nb_H^{++})^2\,, \label{1loop}
 \eea
where the harmonic covariant derivative $\nb_H^{++} = D^{++} +
\alpha_H V^{++}$ depends on the root $\alpha_H$ and $
 \sB_{H} := \square + \a_{H}\, W^{+a} D^-_a \,$.
We also introduced the operators $
 A_{(\pm)}(\Omega) = \Omega \nb_H^{++} \pm \tfrac32 (D^{++}\Omega)$.

The first two terms in the first line of \eq{1loop} are the
contribution from the gauge multiplet and the total contribution
from the hypermultiplet, respectively. The factor
$\mbox{Det}^{1/2}\sB $ in  \eq{Gamma0} produces the first term in the
second line of \eq{1loop}. The last two terms in the second line
come from the ghosts actions.

We divide the one-loop contribution to the effective action \eq{1loop}
into the two terms
 \bea
 \Gamma^{(1)} = \Gamma^{(1)}_{\rm lead} + \Gamma^{(1)}_{\rm high}\,.
 \label{Ghigh}
 \eea
We will see that the first one is responsible for the leading low-energy contribution
 \bea
 \Gamma^{(1)}_{\rm lead} = \f{i}2\Tr_{(2,2)} \ln\Big(\sB_H - \alpha_H^2
 \Omega^2\Big) - \f{i}2\Tr_{(4,0)} \ln\Big(\sB_H -
 \alpha_H^2\Omega^2\Big).  \label{1loop2}
 \eea
As for the second term $\Gamma^{(1)}_{\rm high}$ in \eq{Ghigh}, we will show that it
corresponds to the next-to-leading approximation.
Further we  demonstrate that the $\cN=(1,1)$
invariant action \eq{I1} can be found as a leading
contribution  to the one-loop effective action $\Gamma^{(1)}_{\rm
lead}$ \eq{1loop2}. The action \eq{I1} includes only the gauge
superfield strength $W^{+a}$ and superfield $\Omega$ and does not contain terms
with $D^{++}\Omega, \, D^-_a \Omega$ and $D^-_a W^{+b}$. Hence we will
systematically neglect such terms in our computations. The
contribution $\Gamma^{(1)}_{\rm high}$ collects terms with
$D^{++}\Omega$ and spinorial derivatives of the background
superfields only. Thus, below the contribution
$\Gamma^{(1)}_{\rm high}$ can be ignored.

The scheme of calculation of the contribution \eq{1loop2} is quite  similar to
the analogous one in the four-dimensional case \cite{Kuzenko:2001vc}. First of all
we notice that on shell the harmonic derivative $\nb^{++}_H$ commutes with
the covariant d'Alembertian. But it is not true for the
operator $\sB_H- \alpha_H^2\Omega^2\,$, since $[\sB_H-
\alpha_H^2\Omega^2,\nb_H^{++}]\sim D^{++}\Omega$. However, all such
terms are beyond the scope of our consideration.  Thus, in accordance with the method
of Ref. \cite{Kuzenko:2001vc}, the well-defined expression for the contribution
$\Gamma^{(1)}_{\rm lead}$ to the one-loop effective action reads
 \bea
 \Gamma^{(1)}_{\rm lead}=-\frac{i}{2} \Tr \int_{0}^{\infty}
\frac{ d(is)}{(is)}e^{is(\sB_H- \alpha_H^2\Omega^2)}
\Pi^{(2,2)}_{\rm T} \,. \label{G2_2}
 \eea
Here we have introduced the projection operator on the space of
transverse covariantly analytic superfields, $\Pi^{(2,2)}_{\rm
T}(\zeta_1,u_1; \zeta_2,u_2)$. One can show \cite{Kuzenko:2001vc} that
 \bea
 \Pi^{(2,2)}_{\rm T} = -\f{(D^+_1)^4}{\sB_1}
 \Big\{(\nb^-_1)^4(u^+_1u^+_2)^2
-\Delta_1^{--}(u^-_1u^+_2)(u^+_1u^+_2) + \sB_{1}(u^-_1u^+_2)^2\Big\}
\d^{14}(z_1-z_2)\,, \label{Pi2}
 \eea
where have introduced the notation
$
 \Delta^{--} = i\nb^{ab}\nb^-_a\nb^-_b - W^{-a}\nb^-_a + \frac{1}{4} (\nb^-_a
W^{- a})$. Then we substitute  \eq{Pi2} in the one-loop contribution
$ \Gamma^{(1)}_{\rm lead}$ \eq{G2_2} and take the
coincident-harmonic points limit $u_2 \to u_1$. It is easy to see
that only the third term in (\ref{Pi2}) survives. As the next steps
we collect the terms quartic in the derivative $D^-_a$ from the
exponential in \eq{G2_2} and use the equality $ (D^+)^4 (D^-)^4
\delta^8(\theta_1-\theta_2)\big|_{2=1}~=~1$.  Passing to the
momentum representation and calculating the integral over
proper-time $s$ we obtain
 \bea \Gamma^{(1)}_{\rm lead} =
\f{N-1}{(4\pi)^3}\int d \zeta^{(-4)}\, \f{(W^+)^4}{\Omega^2}\,.
\label{1loop4}
 \eea
The matrix trace in \eq{1loop4} is calculated as a sum
over non-zero roots $\alpha_H$, with $H =
\tfrac{1}{\sqrt{N(N-1)}}{\rm diag}(1,..,1, 1-N)$.

As was expected,  the $\cN=(1,1)$ invariant $I_1$ \eq{I1} comes out as
the leading low-energy contribution \eq{1loop4}  to the effective
action for the theory \eq{S0}. The coefficient $c$ was calculated and it is equal to
 \bea
 c=\f{N-1}{(4\pi)^3}\,. \label{c}
 \eea
It is interesting to note that the same expression for the coefficient $c$ was obtained in
$4D$ $\cN=4$ SYM theory (see, e.g., \cite{Kuzenko:2004sv} and
references therein). The bosonic part of the effective action
\eq{1loop4} is
 \bea
\Gamma^{(1)}_{\rm bos} \sim \int d^6 x\, \f{F^4}{\phi^2}\bigg(1+
\f{\phi^{(ij)}\phi_{(ij)}}{\phi^2}+\ldots\bigg)\,,
 \eea
where $F^4 = 3 F_{MN} F^{MN} F_{PQ} F^{PQ} - 4 F^{NM} F_{MR} F^{RS}
F_{SN}$ and $F_{MN}$ is the abelian gauge field strength.

\section{Conclusion}

Harmonic superspace is a very convenient powerful tool for investigating quantum properties of $6D$ ${\cal N}=(1,0)$
and ${\cal N}=(1,1)$ theories, because it allows to keep ${\cal N}=(1,0)$ supersymmetry manifest at all steps
of calculating quantum corrections. Moreover, this technique considerably simplifies the calculations, because a
huge amount of usual Feynman diagrams appear to be included into an essentially smaller number of superdiagrams.
Surely, most of the statements and methods related to ${\cal N}=(1,0)$ and ${\cal N}=(1,1)$ SYM theories
can be reformulated within the harmonic formalism. The results obtained in the harmonic superspace approach in the lowest loops agree with those found with the help of other techniques,
say, within the component approach. However, the harmonic superspace technique looks certainly more preferable
for calculations in the higher loops, where the advantages of the manifestly supersymmetric quantization method are especially essential.

\section*{Acknowledgements}

The authors are grateful to D.I.Kazakov for valuable discussions and
pointing some references. I.B., E.I. and K.S. acknowledge a partial
support from the RFBR grant, project No 18-02-01046. The work of
I.B., E.I. and B.M was supported by Ministry of Education and
Science of Russian Federation, project No 3.1386.2017.  The work of
B.M. was carried out in part within the Tomsk Polytechnic University
Competitiveness Enhancement Program.


\end{document}